\documentstyle[12pt,epsf]{article}
\begin{document}
\thispagestyle{empty}

\begin{flushright}
DAMTP-95-07 \\
hep-th/9502106
\end{flushright}
\vspace{2.0cm}

\begin{center}

{\bf\Large Multi-boundary effects in Dirichlet String Theory}
\vspace{2.0cm}

{\large Michael Gutperle}
\footnote{e-mail: M.Gutperle@amtp.cam.ac.uk , on leave at
TH-Division, CERN, Geneva} \\
{D.A.M.T.P., Silver Street \\
Cambridge, CB3 9EW, U.\ K.\ }
\vspace{0.5cm}

\end{center}

\begin{abstract}
String amplitudes with an arbitrary number of world-sheet boundaries
on which the coordinates satisfy Dirichlet boundary conditions are
analyzed in a path integral framework.  Special attention is payed to
the novel divergences associated with such conditions.  Certain
helicity amplitudes involving massless closed-string states are free
of such divergences to all orders in perturbation theory and their
behavior can be analysed unambiguously.  The high energy fixed-angle
behavior of these amplitudes is discussed in the presence of one or
two boundaries and the asymptotic behavior of the amplitudes is shown
to be power behaved.
\end{abstract}

\section{Introduction}
In open string theories one has to impose boundary conditions on the
fields living on the world sheet in order to canonically quantize the
theory or to define a path integral. For the usual bosonic open
string one imposes Neumann conditions on the space-time coordinates
$X^{\mu}$, which means that  the normal derivatives of $X^{\mu}$ are
set equal to zero at the boundary. Another possibility consistent
with conformal invariance \cite{cardy} is to impose constant
(Dirichlet) boundary conditions. This means that each boundary is
mapped into a single spacetime point.

This idea was first employed to formulate off-shell extensions of
dual amplitudes by coupling dual models to external currents, for
open strings \cite{schwarz}, \cite{corrigan} and for  closed strings
\cite{mbgd}. See also \cite{cohen} for a path integral approach to
off-shell amplitudes using Dirichlet boundary conditions .  The
theory with Dirichlet boundaries is formally related to the Neumann
theory by target space duality of the compactified theory
\cite{mbgtsd},\cite{pol2}, but this will not be considered here since
we will be concerned with flat Minkowski space as a target space
manifold.

This paper will investigate the theory with Dirichlet boundaries in
which the boundary positions are integrated over space-time
\cite{mbgtsd}. This  ensures that no momentum can enter or leave
through any boundary leading to a modification of the closed string
theory. The insertion of a single Dirichlet boundary was considered
in \cite{mbgtsd} and the case of two boundary insertions was
considered in \cite{li}, \cite{zhang}. The theory has drastically
altered fixed angle scattering amplitudes for closed string states
which exhibit pointlike behavior in the high energy fixed angle
regime.
This paper will discuss properties of the string perturbation
expansion for such a theory on oriented Riemann surfaces although the
doubling procedure that we use can also be applied to unoriented
surfaces and many features carry over to Dirichlet boundary
insertions on nonoriented Riemann surfaces.

The organization of the paper is as follows: In section \ref{general}
we describe the Dirichlet open string by doubling the bordered world
sheet and all the quantities which are necessary for the evaluation
of the open string path integral are expressed in terms of quantities
on the compact double. The  path integral with an arbitrary number of
world-sheet boundaries on which the coordinates satisfy Dirichlet
boundary conditions is performed and different applications for the
result are given. Certain  new kinds of divergences  arise in the
Dirichlet theory that originate from the two lowest states ($N=0$,
$N=1$) in the open string Hilbert space. A general analysis of these
divergences and of the degenerations of the world sheet where they
occur is given.
In section \ref{one} and \ref{two} we analyze the scattering
amplitudes with one and two  Dirichlet boundary insertions. The
analysis of the divergences of section \ref{general} is illustrated
in these two cases.  We focus on  high energy fixed angle scattering
and show that for certain helicity amplitudes of massless tensor
states the behavior of the amplitudes can be analyzed unambiguously.
The scattering behavior is disentangled from the divergences which
arise at the boundary of moduli space and it is shown that the
amplitudes are power behaved with respect to the center of mass
energy.

In appendix A  a modification of the Green function is discussed
which arises in the functional integration. The r\^ole of the $N=1$
state and its interpretation as a Lagrange multiplier field will be
described in Appendix B. In a scheme due to  Polchinski \cite{pol3}
the combinatorics of the boundary insertions are altered and this
leads to a cancellation of the divergence caused by this field. This
mechanism is briefly illustrated in Appendix C.

\section{General Considerations}

\label{general}
The action of the bosonic string in flat space time is given by
\begin{equation}
S =\frac{1}{4\pi \alpha^{\prime}} \int_{\Sigma}{d^2\sigma \sqrt{detg}
g^{ab}\partial _a X^{\mu} \partial _b X_{\mu}}
\label{eq:action}
\end{equation}
Scattering amplitudes for closed bosonic strings are defined
perturbatively by the Polyakov-path-integral over surfaces of
different topologies and given by expectation values of
vertex-operators inserted on the world sheet
\begin{equation}
\Gamma = \sum_{topologies}\int{\frac{dg dX}{vol(diff\lhd
weyl)}\exp(-S(X,    g)) \prod_{i=1}^n \int {d^2 \sigma_i V(k_i,
X(\sigma_i))}}
\label{path1}
\end{equation}
Holomorphic factorization \cite{knizh} of the path integral measure
implies that the vacuum amplitude with no vertex operator insertions
can be written as an integral over the complex moduli space and the
integrand is given by a modular form on this space
\begin{equation}
Z_g=\int_{moduli}\prod_{k=1}^{3g-3}d^2m_k | \sigma(m_1,    \cdots,
m_k)|^2\det(Im(\tau))^{-13}
\label{eq:measure}
\end{equation}
Here $\prod_{k=1}^{3g-3}d^2m_k|\sigma(m_1,    \cdots,    m_k)|^2$ is
due to the evaluation of the path integral and contains the
determinants of scalar and vector laplacian of the Riemann surface,
$\sigma(m)$ is a modular form of weight $-13$ on the moduli space of
complex dimension $3g-3$. It is parameterized by the coordinates
$m_i$ which correspond to the deformations of the complex structure
of the surface. The measure $\sigma(m)$ can be expressed in terms of
well known functionals which are defined on the Riemann surface
corresponding to the moduli $m$  \cite{verlinde}.
The measure can also be represented by a path integral over ghosts in
the usual way and one gets
\begin{equation}
Z_g=\int_{moduli}\prod_{k=1}^{3g-3}d^2m_k \int DX Db Dc
\prod_{i=1}^{3g-3}|(\mu_i,b)|^2 e^{-S_{gh}-S_X}
\label{eq:ghost}
\end{equation}
Here the $(\mu_i,b)$ is the pairing of a Beltrami differential
corresponding to the modulus $m$ with a b-ghost to absorb the zero
modes of the b-ghost. This formula is valid for $g>1$, for low genus
one also has to take care of the zero modes for the c-ghosts
corresponding to conformal Killing vectors.

On a compact Riemann surface we can define an intersection pairing
$J$ and a canonical homology basis $(a_i,    b_i),    i=1,    \cdots,
   g$ which satisfies
\begin{equation}
J(a_i,    a_j)=J(b_i,    b_j)=0 \mbox{ ; }
J(a_i,    b_j)=-J(b_j,    a_i)=\delta_{ij}
\label{eq:inters}
\end{equation}
Together with the homology cycles come the Abelian differentials  of
the first kind $\omega_i$ ,    which define the period matrix $\tau$
by
\begin{equation}
\oint_{a_i}\omega_j=\delta_{ij}  \mbox{ ; }
\oint_{b_i}\omega_j=\tau_{ij}
\label{eq:diff}
\end{equation}


\subsection{Doubling of a bordered surface}
In mathematics \cite{shiff},    conformal field theory \cite{cardy}
and open string theory one encounters Riemann surfaces which have
boundaries,    and therefore are not compact. A standard way of
dealing with this is to represent the bordered Riemann surface as a
quotient of a compact Riemann surface under an antiholomorphic
involution. This is called doubling and has been used to analyze the
loop amplitudes of the standard (Neumann) open string theories
\cite{burgess},    \cite{blau}. Using this the  well known results
for closed strings can also be applied in the context of Dirichlet
boundary conditions.

Starting with a (oriented) surface $\Sigma$ with $b$ boundaries
$\partial \Sigma_i,    i=1,    \cdots ,    b$ and $h$ handles, we
take a topological copy of $\Sigma$ and glue the two surfaces
together by an orientation reversing involution $I$ such that
$\Sigma=\overline{\Sigma}/I$ and
$I(\partial\Sigma_i)=\partial\Sigma_i$ (See figure 1).  A complex
structure on $ \overline{\Sigma}$ can be defined from one on $\Sigma$
by applying the Schwartz reflection principle : For a local complex
coordinate $z(p)$ near $p\epsilon\Sigma$ a coordinate at $I(p)$ is
given  via $z(I(p))=\overline{z}(p)$. This makes $\overline{\Sigma}$
into a complex manifold and the involution $I$ anticonformal. The
double is a compact surface of genus $g=2h+b-1$.
In general the involution $I$ acts nontrivially on the homology
cycles $a_i,    b_i$;  $ 1,    \cdots,    g$
\begin{eqnarray}
I(a_i) & = & \sum_{j} C_{ij} a_j \\
I(b_i) & = & \sum_{j} D_{ij}a_j - \sum_j C_{ij}b_j
\end{eqnarray}
With certain constraints on the matrices $C,    D$ coming from the
invariance of (\ref{eq:inters}) under $I$. A convenient choice of the
homology cycles which is possible \cite{sagnotti} is given by
$D_{ij}=\delta_{ij}$. In the most general cases one cannot set
$C_{ij}=0$, but when there are no handles, i.e. $h=0$, this can be
done. The invariance of the a-cycles means that they either lie on
the boundary or are sums of cycles which get interchanged by $I$. An
important term which appears in the expressions of the functional
determinants is
\begin{equation}
R_{\Sigma}=det\{\frac{1}{2}(1-D)Im(\tau)^{-1}
+\frac{1}{2}(1+D)Im(\tau)\}
\end{equation}
With the special choice above this gives $R_{\Sigma}=Im(\tau)^{-1}$.
In \cite{blau} the functional determinants which occur during the
reduction of the path integral to an integral over moduli space for a
bordered surface were obtained from the determinants for the compact
double. The measure on moduli space of the open string is given by
\begin{equation}
\prod_{k=1}^{3g-3}dm_k|\sigma(m_1,    \cdots,    m_k)|[\det
Im(\tau)]^{-13/2}R_{\Sigma}^{\stackrel{+}{-}13/2}
\label{eq:measure2}
\end{equation}
In principle this is the `holomorphic square root' of
(\ref{eq:measure}) and is to be integrated over a real slice of the
moduli space of the genus g surface. The involution $I$ also acts on
the moduli space and in essence the part of moduli space which is
invariant under $I$ (those deformations which deform both $\Sigma$
and $I(\Sigma)$ symmetrically) is integrated over. The sign in
(\ref{eq:measure2}) is determined by the boundary condition on the
spacetime fields $X$,     plus for Neumann and minus for Dirichlet
boundary conditions, see \cite{blau} for details.

\subsection{Gaussian integration}
For a surface with a boundary the integration by parts in the action
(\ref{eq:action}) gives a boundary term contribution, we drop the
antiholomorphic arguments for notational convenience.
The most general vertex operator has the form $V(k,\zeta)=:A(\partial
X,. . ,    \zeta)e^{ikX}:$ , where A symbolically indicates the
dependence on the polarization tensors $\zeta$ and the derivatives of
$X$ and the ellipses denote normal ordering. The functional integral
with a product of such vertex operators can be evaluated by
exponentiating the prefactors $A$. This leads to terms involving the
derivatives of the Green function but without any momentum
dependence. Since we focus on the momentum dependence of the
amplitudes leaving the prefactors unexponentiated and combining the
universal factors $\exp(ikX)$ gives the exponential part of the
integrand $\epsilon$
 \begin{equation}
\epsilon=-\frac{1}{4\pi\alpha^{\prime}}\int_{\Sigma}{X
\partial \overline\partial X} +
\frac{1}{4\pi\alpha^{\prime}}\int_{\partial\Sigma}{X\partial
X}+i\sum_{i=1}^{n}k_i^{\mu}X_{\mu}(z_i)
\label{eq:byparts}
\end{equation}
To integrate out the $X$ in the path integral (\ref{path1}) $X$ is
separated into a classical part and a quantum fluctuation $\xi(z)$
\begin{equation}
X^{\mu}(z) = \xi^{\mu} (z) +2\pi i\alpha^{\prime}\sum_{i=1}^n
k_i^{\mu}G^D(z,    z_i) +
X_{cl}^{\mu}(z)
\label{eq:classical1}
\end{equation}
 $G^D (z,    w)$ is the (uniquely defined) Dirichlet Green function
on the
surface $\Sigma$ with $b$ boundary components $\partial \Sigma_i$,
 $i=1,    \cdots,    b$ which vanishes on all boundary components.
The quantum fluctuation $\xi$ is also chosen to vanish at the
boundaries.
The classical solution $X_{cl}(z)$ satisfies the Laplace-equation in
$\Sigma$ and satisfies on the boundary
$X_{cl}(z)|_{z\epsilon\partial\Sigma_i}=Y_i$.
 It is easy to see that $X_{cl}$ can be expressed implicitly with the
help of the
Green function in the following way
\begin{equation}
X_{cl}(z)=2i\sum_{i=1}^b\oint_{\partial\Sigma_i}
{dwX_{cl}(w)\partial_w
G^D (z,w)}
\label{eq:classical2}
\end{equation}
The first $b$ a-cycles are chosen to lie on the boundary and the
boundaries are represented by
\begin{equation}
\partial \Sigma_1 = a_1,     \partial \Sigma_2 = a_2 -a_1 ,
\cdots,    \partial
\Sigma_{b-1} = a_{b-1} - a_{b-2},     \partial\Sigma_b
=\partial\Sigma-a_{b-1}
\label{eq:homology}
\end{equation}
With this definition  the $X_{cl}$ can be expressed as an integral
over the a-cycles and the boundary values $Y_i$
\begin{equation}
X_{cl}(z)=2i\sum_{i=1}^{b-1}(Y_i-Y_{i+1})\oint_{a_i}{dw\partial_w
G^D (z,    w)}+Y_b
\label{eq:classical3}
\end{equation}
The Green function on the Riemann surface $\Sigma$ $G^D_{\Sigma}(z,
 w)$
can be given in terms of the Green function on the double
$\overline{\Sigma}$ which in turn can be expressed in terms of the
prime-form $\it E$ and abelian differentials of the first kind
$\omega_i$ \cite{dhoker}
\begin{equation}
G^D_{\Sigma}(z,    w) = G_{\overline{\Sigma}}(z,    w)
-G_{\overline{\Sigma}}(z,    I(w))
\label{eq:greensf1}
\end{equation}
\begin{equation}
G_{\overline{\Sigma}}=
-\frac{1}{4\pi}\ln|E(z,    w)|^2+\frac{1}{2}Im\int^w_z\omega_i
Im\tau_{ij}^{-1}Im\int^w_z\omega_j+F(z,    \overline{z})+F(w,
\overline{w})
\label{eq:greensf2}
\end{equation}
The Gaussian integration can be done by inserting
(\ref{eq:classical1}) into (\ref{eq:byparts}). Omitting a term
quadratic in the quantum fluctuations $\xi$, which gives the
determinant of the Laplacian upon functional integration of $\xi$,
the classical contribution to the exponent $\epsilon_{cl}$ is
\begin{equation}
 \epsilon_{cl} = -\alpha^{\prime}\pi\sum_{ij}{k_i k_j
G^D_{\Sigma}(z_i ,    z_j
)}+\frac{1}{4\pi\alpha^{\prime}}\int_{\partial\Sigma}{X_{cl}\partial
X_{cl}}+ i\sum_ik_i X_{cl}(z_i )
\label{eq:gauss1}
\end{equation}
 Using (\ref{eq:classical2}) and (\ref{eq:greensf2}) the double
derivative of the Green function is given by
\begin{equation}
\partial_z\partial_w
G^D(z,    w)=-1/(4\pi)\omega(z,    w)-1/2\sum_{i,
j}\omega_i(z)Im\tau_{ij}^{-1}\omega_j(w)
\end{equation}
 Here $\omega(z,    w)$ is the abelian differential of the third kind
which vanishes when integrated along a-cycles. The second term on the
r.h.s. of (\ref{eq:gauss1}) can then be evaluated with the help of
(\ref{eq:diff}).
\begin{eqnarray}
\int_{\partial\Sigma}X_{cl}\partial X_{cl} & = &
+2i\sum_{ij}(Y_i-Y_{i+1})(Y_j-Y_{j+1})\oint_{a_i}\oint_{a_j}\partial_
z\partial_w G^D(z,    w) \nonumber \\
&=& -i\sum_{i,    j} (Y_i
-Y_{i+1}) Im\tau_{ij}^{-1} (Y_j - Y_{j+1})
\label{eq:dxcl}
\end{eqnarray}
This term and the third term on the r.h.s. of (\ref{eq:gauss1}) give
the explicit dependence of the amplitude  on the boundary values
$Y_i$. There are several uses for these formulas:

$\bullet$ If there are no vertex operator insertions (\ref{eq:dxcl})
is the only contribution in $\epsilon_{cl}$ from the Gaussian
integration. We can interpret these amplitudes which depend on
$Y_i-Y_{i+1}$ as space time correlation functions.
These functions have interesting singularity structure with
singularities outside,    on and inside the lightcone \cite{mbgtsd}.
Putting together (\ref{eq:measure2}) and (\ref{eq:dxcl}) gives with
$\Delta Y_i=Y_i-Y_{i+1}$ for the special case of no handles, i.e.
$h=0$
\begin{equation}
A(\Delta Y)=\int\prod_{k=1}^{3g-3}dm_k |\sigma(m)| \det
Im(\tau)^{-13}\exp(\frac{1}{2\pi\alpha^{\prime}}\sum_{i,    j}\Delta
Y_i Im(\tau_{ij})^{-1}\Delta Y_j)
\label{eq:partition}
\end{equation}
Note that the $\Delta Y_i$ enter this expression just like the loop
momenta enter the expression for the partition function in the
Neumann theory.  Integrating over the $\Delta Y_i$ gives
\begin{equation}
\int\prod_{i=1}^{b-1}d^{26}\Delta Y_i A(\Delta
Y)=(2\pi\alpha^{\prime})^{13}\int\prod_{k=1}^{3g-3}dm_k |\sigma(m)|
\end{equation}
 This is the same expression as the  Neumann partition function given
by (\ref{eq:measure2}) identifying the loop momenta $P_i$ with
$\Delta Y_i/(\sqrt{2\pi\alpha^{\prime}})$

\bigskip
$\bullet$ Inserting closed string vertex operators gives amplitudes
which  have been interpreted as coupling of currents to the
(hadronic) string \cite{mbg2}. One obvious thing to consider is a
Fourier transformation with respect to the boundary positions. This
gives off-shell amplitudes \cite{mbg2},    \cite{cohen}. Starting
from (\ref{eq:gauss1}) it is easy to see that $\epsilon_{cl}$ only
depends on the $\Delta Y_i$and $Y_b$. Defining momenta $q_i$
conjugate to $Y_i$  an overall momentum conserving delta function
$\delta(\sum_i q_i+\sum_jk_j)$ is obtained from the integration of
$Y_b$ and an integrand which then only depends on the $\Delta Y_i$.
With new momenta $\hat q_i$ conjugate to the $\Delta Y_i$
\[
A(k_i, \hat q_j)=\int\prod_{k=1}^{3g-3}dm_k |\sigma(m)| \det
Im(\tau)^{-13}
  \int\prod_{k=1}^{b-1} d^{26}\Delta Y_k \exp(i\sum_{k=1}^{b-1}\hat
q_k \Delta Y_k +\epsilon_{cl}) \nonumber
\]
\begin{equation}
=\int\prod_{k=1}^{3g-3}dm_k
|\sigma(m)|\exp\{-2\pi\alpha^{\prime}\sum_{l,    m=1}^{b-1}(b-i\hat
q)_lIm(\tau)_{lm}(b-i\hat q)_m\}
\label{eq:dirich1}
\end{equation}
Where $b_j=\sum_ik_i\oint_{a_j}\partial_w G^D(z_i,    w)$. Note that
the Fourier transformation has changed the powers of $\det(Im(\tau))$
in the measure.  The $\hat q_i$ are connected to the momentum of the
Dirichlet boundaries and one interesting thing to consider is the
`deep inelastic' regime where $|q_i|\to\infty$ corresponding to a
large momentum transfer through the boundaries into the diagram.

\bigskip
$\bullet$ Another interesting consideration is the situation when all
the $Y_i$ are integrated independently over space time with unit
weight, so all $\hat q_i$ are set to zero. This means that no
momentum is allowed to flow through  the boundaries. Setting  $\hat
q_i=0$ in (\ref{eq:dirich1}) produces a modification of the Green
function since $\epsilon_{cl}=\pi\alpha^{\prime}\sum k_i k_j
\hat{G}(z_i,    z_j)$ with
\begin{equation}
\hat{G}^D(z,    w) =
G^D(z_i,    z_j)-2\sum_{l,    m}\oint_{a_l}\partial{w}G^D(z_i,
w)Im(\tau)_{lm}\oint_{a_m}\partial_{w^{\prime}}G^D
(z_j,    w^{\prime})
\label{eq:modgreen1}
\end{equation}
 In appendix A it is  shown that the second term on the right hand
side of (\ref{eq:modgreen1}) cancels the second term on the right
hand side of (\ref{eq:greensf2}) and the modified Green function is
given by
\begin{equation}
\hat{G}^D(z_i,    z_j)=-\frac{1}{4\pi}\{ln\left|E(z_i,
z_j)\right|^2-ln\left|E(z_i,    I(z_j))\right|^2\}
\end{equation}
The resulting  modified closed string theory  has very different
short distance properties compared to ordinary closed string theory.
For example the high temperature properties \cite{mbg4} of this
theory are similar to a QCD like theory \cite{pol1}. Another probe of
the short distance properties of a theory is high energy fixed angle
scattering. For two particle scattering amplitudes this is the limit
$s\to\infty$ while $s/t$ is fixed (s,t are Mandelstam variables
defined by $s=-(k_1+k_2)^2,t=-(k_1+k_3)^2$). In general one considers
$|k_i k_j|\to\infty$ and $|k_ik_j|/|k_kk_l|$ fixed. The absence of
any short distance structure in conventional closed string theories
is reflected by the exponential decrease of fixed angle scattering as
a function of $s$ which is the center of mass energy. This was
already noted in the early days of string theory
\cite{veneziano},\cite{amati}. In \cite{gross1} it was shown that the
amplitude in this limit is dominated by a saddle point configuration
and the high energy fixed angle behavior is universal to all orders
in the topological expansion of perturbative string theory. Here the
situation is different   \cite{mbgtsd}, since there is  a power law
fall off in $s$,    which can be interpreted as  pointlike scattering
behavior induced by the Dirichlet boundary insertion compared to the
complete lack of pointlike structure in the ordinary closed string.
Because of the structure of the Green function in (\ref{eq:greensf1})
we can represent a configuration with $n$ Vertex operators $V_i(z_i)$
with momentum $k_i$ by $2n$ Vertex operators where
$V_{n+i}$ is located at $I(z_i)$, which is the `mirror image' of
$z_i$, and has momentum $-k_i$. Then $\epsilon_{cl}$ clearly vanishes
 when all $z_i=I(z_i)$,   i.e. when all the $z_i$ are on the
boundary. In an electrostatic picture the $z_i$ and $I(z_i)$ are the
positions of charges on the surface with strength $k_i^{\mu}$ and
$-k_i^{\mu}$. It is clear that in the configuration above the charges
annihilate each other and the energy $\epsilon_{cl}$ vanishes. The
asymptotics are dominated by an endpoint integration. This means that
the integration over the deviations $\delta z_i$ gives powers of of
the kinematical invariants. As explained later we expect the $G^{D}$
to be quadratic in the deviations and the generic behavior should be
$s^{-n/2}$ where $n$ is the number of vertices. This naive picture is
complicated by the fact that  the dominant region of the integrand
lies on the boundary of moduli space,    where we encounter new kinds
of divergences. In the following sections the relevant divergences of
the Dirichlet string are discussed  and they are dealt with in the
easiest cases.

\bigskip
$\bullet$ A different version of Dirichlet string theory was proposed
by Polchinski  \cite{pol2}. This is defined by considering a gas of
`D-instantons'. A single D-instanton is defined at a point $Y_{\mu}$
in space time, and is build up by an arbitrary number of disconnected
(in the topological sense) world sheets, which all have the
boundaries mapped to $Y$. The disconnected world sheets are really
connected in space time via the common point Y. We can easily adapt
our formalism to this description and it means that for each
connected world sheet one has to set $\Delta Y_i=0$, for all $i$ and
$Y_b=Y$. So that in  (\ref{eq:modgreen1}) the second term on the
r.h.s. is not present. The full theory is then given by summing over
an arbitrary number of such D-instantons and integrating over their
space time positions.


\subsection{Divergences in Dirichlet amplitudes}
\label{divergences}
It is well known that infinities in string amplitudes arise from the
boundary of moduli space. In closed string theory these boundaries
originate either when vertex operators approach each other on the
world sheet or when surfaces degenerate and a cycle is pinched. In
the plumbing fixture construction one represents this as a cylinder
glued into the surface which becomes infinitely long \cite {dhoker}.
The standard divergences that arise in closed bosonic string theory
will not be discussed here since we are interested in new features
that are due to the presence of Dirichlet boundaries.

 In open string theory there are four other types of  degenerations
(see figure 2):
(a) vertex operator positions approaching the boundaries,    (b) the
separation of the surface into two disconnected parts,    (c) two
boundary components touching,    (d) a boundary shrinking to a point.
(d) corresponds to a closed string tadpole coupling to the disk,    a
phenomenon which is well discussed in the literature for Neumann
strings \cite{callan1}. We do not discuss the case (d) in any detail
here since our main focus is on the novel open string divergences in
the theory.
The degenerations (a) to (c) can be represented in a conformal frame
where a strip glued to the boundaries of the surfaces becomes
infinitely long (see figure 3). This degenerating strip can be
represented by vertex operator insertions for all open string states
at the two points where the strip leaves the surface and a propagator
joining them. Symbolically this can be represented as
\begin{eqnarray}
A_{(a)} &=& \sum_{\Phi_1,    \Phi_2}\langle V_1. .
V_{n-1}A_{\Phi_1}\rangle_{\Sigma_1}\langle\Phi_1|\Delta|\Phi_2\rangle
\langle V_n A_{\Phi_2}\rangle_{Disk} \nonumber \\
A_{(b)} &=& \sum_{\Phi_1,    \Phi_2}\langle V_1. .
V_{n-l}A_{\Phi_1}\rangle_{\Sigma_1}\langle\Phi_1|\Delta|\Phi_2\rangle
\langle V_{n-l+1}. . V_n A_{\Phi_2}\rangle_{\Sigma_2} \nonumber \\
A_{(c)} &=& \sum_{\Phi_1,    \Phi_2}\langle V_1. .
V_{n}A_{\Phi_1}A_{\Phi_2}\rangle_{\Sigma^{\prime}}
\langle\Phi_1|\Delta|\Phi_2\rangle \nonumber \\
A_{(d)} &=& \sum_{\hat\Phi_1,    \hat\Phi_2}\langle V_1. .
V_{n}A_{\hat\Phi_1}\rangle_{\Sigma^{\prime}}\langle\hat\Phi_1|
\hat\Delta|\hat\Phi_2\rangle \langle A_{\hat\Phi_2}\rangle_{D}
\label{eq:sewing}
\end{eqnarray}
Here $A_{\Phi_i}$ denotes a boundary vertex operator insertion
corresponding to the state $\Phi_i$ in the open string Hilbert space
and $\Delta$ is the open string propagator,    the hatted quantities
$\hat\phi$ and $\hat\Delta$ denote closed string states and
propagator for case (d).  The functional integral over $\Sigma$ is
denoted by $\langle. . \rangle_{\Sigma}$.
In the case of Dirichlet boundaries consider a free string quantized
on an infinite strip of width $\pi$ with
\begin{equation}
X^{\mu}(\sigma,    \tau)|_{\sigma=0}=Y_1^{\mu} \mbox{ , }
X^{\mu}(\sigma,    \tau)|_{\sigma=\pi}=Y_2^{\mu}
\label{eq:dirquant}
\end{equation}
The mode expansion of the string coordinate $X$ is then given by
\begin{equation}
X^{\mu}(\sigma,\tau)=Y_1^{\mu}+\frac{Y_2^{\mu}-Y_1^{\mu}}
{\pi}\sigma+\sqrt{2\alpha^{\prime}}\sum_{n+-\infty}^{n=\infty}\frac{1}{n
}
a_n^{\mu}\sin
n\sigma e^{in\tau}
\end{equation}
The Virasoro generators $L_n$ are easily obtained from this expansion
and the zero mode of the energy momentum tensor $L_0$ is given by
\begin{equation}
L_0^{(Y_1-Y_2)}=\frac{(Y_1-Y_2)^2}{4\pi\alpha^{\prime}}+N
\label{eq:L0}
\end{equation}
N is the open string number operator given by
$N=\frac{1}{2}:\sum_{n}a_{-n}^{\mu}a_{n\mu}:$. The propagator is
given by $\Delta = (L_0-1)^{-1}$ and can be written in an integral
representation in the following way
\begin{equation}
\Delta=\int_{0}^{\infty}dt e^{-t(L_0-1)}
\label{eq:propagator}
\end{equation}
In this representation it is clear that the divergences in the
degenerations (a)-(c) arise from the limit $t\to\infty$ due to the
two lowest lying states namely the $N=0$ and the $N=1$ state, when
the boundary values $Y_1$ and $Y_2$ coincide.
For the $N=0$ state this is very similar to the usual tachyon
divergence at zero momentum in Neumann theory. For the $N=1$ state
the propagator $\Delta$ is infinite if $Y_1=Y_2$, which is the novel
divergence in Dirichlet string theory.
Note that while in (a) and (b) the degeneration automatically
enforces $Y_1=Y_2$ on the degenerating strips since there is only one
boundary component degenerating,    this is not the case for (c)
since there are now two boundaries.  In  Polchinski's scheme these
are automatically set equal whereas we will integrate over $Y_1$ and
$Y_2$. The above mentioned position space singularities in
$(Y_i-Y_{j})^2$ come  from the (c) degeneration as shall  be seen
later.

To get a more explicit picture the degenerations (a)-(c) can be
represented as degenerations on the compact double $\overline\Sigma$
(see figure 4) and  the well known properties of closed string
integrands on the boundary of moduli space \cite{verlinde} can be
used .
To discuss the case (a) we choose $z_n\to \partial\Sigma_1$ where the
boundary is at the space time point $Y_1$ and a local coordinate
system where $I(z_n)=\overline{z}_n$ and $\partial\Sigma_1$ is
represented by the real line. For simplicity  consider ground state
scattering amplitudes, this will result in an expression which is a
special case of $A_{(a)}$ in (\ref{eq:sewing}). Writing $z_n=x+i\eta$
the divergence in the limit $\eta\to 0$ comes from the self
contraction $:G^D(z_n,    z_n):$ in $\epsilon_{cl}$. In the path
integral normal ordering is equivalent to subtracting out the
logarithmic divergence
\begin{equation}
:G^D(z_n,    z_n):=\lim_{z\to z_n}\{G^D(z,
z_n)-\frac{1}{4\pi}\ln|z-z_n|^2\}
\end{equation}
If the arguments of the prime form in (\ref{eq:greensf2}) are close
we have
\begin{equation}
E(z,w) = (z-w) + S(z)(z-w)^3 +o(|z-w|^4)
\end{equation}
Here $S(z)$ is the projective connection on $\overline\Sigma$
\cite{Fay}. The limit $\eta\to 0$ gives  $:G(z_n,
z_n):=\frac{1}{2\pi}\ln|\eta |[1+o(\eta^2)]$ since all other
contributions in (\ref{eq:greensf2}) are of order $\eta^2$.
Setting $k_n^2=4/\alpha^{\prime}$ for closed string tachyons the
divergence has the form $\exp(\epsilon_{cl})\sim \eta^{-2}$. This
divergence is due to the level $N=0$ state. To get the contribution
from the level $N=1$ state the terms linear in $\eta$ in
$\epsilon_{cl}$ have to be extracted.  For $i\neq n$ the Green
function vanishes when one of the arguments is at the boundary which
means that for small $\eta$
\begin{equation}
G(z_n,    z_i)= i\eta\partial_{\eta}G^D(x+i\eta,
z_i)|_{\eta=0}+o(\eta^2)
\end{equation}
Therefore the term linear in $\eta$ coming from the exponent is
\begin{equation}
i\sum_{i=1}^{n-1}k_nk_i \partial_{\eta}G^D(z_n,
z_i)|_{z_i\epsilon\partial\Sigma_1}
\end{equation}
Note that the derivative with respect to  $\eta$ is the normal
derivative with respect to  the boundary $\partial \Sigma_1$. The
Disk amplitude $\langle V(k_n)\oint\bf{1}\rm\rangle _{disk}$ with one
closed string tachyon vertex operator $V(k_n)$ and a $N=0$ boundary
vertex operator $\oint\bf{1}\rm$  is given by a phase factor
$\exp(ik_nY_1)$ which is important for overall momentum conservation.
The Disk diagram with one closed string tachyon operator and one
insertion of a $N=1$ boundary vertex operator$\oint\partial_n X$ is
given by $\langle V(k_n)\oint\partial_nX^{\mu}\rangle_{disk}$ and is
proportional to $k_n^{\mu}\exp(ik_nY_1)$. On the other part of the
surface $\Sigma$ the insertion of $n-1$ closed string tachyon vertex
operators is $\langle \prod_{i=1}^{n-1}V(k_i)\oint\partial _n
X^{\mu}\rangle_{\Sigma}=\sum_{i=1}^{n-1}k_i^{\mu}\oint dz\partial_n
G^D(z,    z_i)$, where $\langle \rangle_{\Sigma}$ denotes the
expectation value on the surface $\Sigma$. Putting this all together
the divergent part of the  amplitude is given by
\[
A_{div}^{(a)}= \int d\eta\eta^{-2}\langle V_1. . V_{n-1}\oint\bf
1\rm\rangle_\Sigma \langle \oint\bf 1 \rm V_n\rangle_D
\]
\begin{equation}
 +\int d\eta \eta^{-1}\langle V_1. .
V_{n-1}\oint\partial_nX_{\mu}\rangle_\Sigma \langle
\oint\partial_nX^{\mu} V_n\rangle_D
\label{eq:limit}
\end{equation}
Note that the level 0 state is just the (integrated) tachyon operator
which is inserted in the first term in (\ref{eq:limit}).
The $N=0$ divergence is very much like a tachyonic divergence at zero
momentum which one encounters in closed string theory. It has been
argued \cite{marcus}
that such divergences are due to the failure of the integral
representation for the propagator in (\ref{eq:propagator}) and that
the  integrals can be defined by analytic continuation,
\begin{equation}
\int \frac{dl}{l^2}=\lim_{s\to 0}\int
\frac{dl}{l^{2+s}}=-\frac{1}{1+s}=-1
\end{equation}
Although subject to ambiguities, this argument supports the
conjecture that only really dangerous divergences are logarithmic
divergences like $\int dl/l$.
In the context of the scattering of massless states the $N=0$
divergence is not there if one considers special helicity amplitudes
as shall be seen in the next section.

To discuss the cases (b) and (c) the plumbing fixture
parameterization has to be modified  in a way which is compatible
with the involution $I$.
The case (b) corresponds to a pinched zero homology cycle  and the
case (c) to a pinched b-cycle. The pinching of an a-cycle corresponds
to case (d).
The construction of the plumbing fixture is as follows
\cite{verlinde}.  Choose two points $p_i$ with $i=1,2$ and coordinate
disks $D_i=\{|z_i|<1\}$ on two disconnected surfaces $\Sigma^{(i)}$
for (b) or on one surface $\Sigma$ for (c). For a fixed complex $t$
with $0<|t|<1$  remove smaller disks $D_i^t=\{|z_i|<|t|^{1/2}\}$ and
glue the remaining surfaces and a cylinder
$C_t=\{w:|t|^{1/2}<w<|t|^{-1/2}\}$ together in the following fashion:
\begin{equation} w = \left\{
\begin{array}{ccc}
\frac{t^{\frac{1}{2}}}{z_1} & \mbox{for} & |t|^{\frac{1}{2}}<|w|<1
\\
\\
\frac{z_2}{t^{\frac{1}{2}}} & \mbox{for} & 1<|w|<|t|^{-\frac{1}{2}}
\\
\end{array}
\right.
\end{equation}
The appropriate coordinates on the moduli space are $d^2p_1$,
$d^2p_2$,    $d^2 t$ and the coordinates of the moduli space of the
two separate surfaces. For the open string the moduli space has to be
restricted to a real slice. This can be done choosing $t$ to be real
and the $p_i$ to lie in the appropriate boundary component,   i.e.
the $p_i$ are invariant under $I$. The integration volume element is
then given by $dt$,    $dp_1$,    $dp_2$ where this implies a
integration along the boundary components.

To discuss the case (b) behavior of the terms in (\ref{eq:measure2})
and (\ref{eq:gauss1}) under $t\to 0$ is important, adopting the
notation that an index $ ^{(i)}$ denotes evaluation on the
degenerated surface $\Sigma_i$. In \cite{verlinde} it is shown that
closed string measure (\ref{eq:measure}) factorizes into two parts
depending on the modular parameters on $\Sigma_1$ and $\Sigma_2$ and
the divergent part $d^2t |t|^{-4}$. The measure on the bordered
surface is given by (\ref{eq:measure2}) and we see that the divergent
part is given by $dt/t^{-2}$. To fix notation we assume that
$\Sigma_1$ contains the first $k$ homology cycles $(a_i,    b_i):i=1,
   . . ,    k$ and that $\Sigma_2$ contains the rest $(a_1,
b_i):i=k+1,    . . ,    g$.   For a holomorphic differential on the
first degenerated surface, i.e. $i\epsilon\{1,    . . ,    k\}$ the
limiting behavior as  $t\to 0$ is given by
\begin{equation}
\omega_i(z) \to \left \{
\begin{array}{ccc}
 \omega_i^{(1)}(z)
+\frac{1}{4}t\omega^{(1)}(p_1)\omega_{p_1}^{(1)}(z),     & \mbox{for}
& z\epsilon \Sigma^{(1)} \\
\\
 \frac{1}{4}t\omega^{(1)}(p_1)\omega_{p_2}^{(2)}(z) + O(t^2) &
\mbox{for} & z\epsilon \Sigma^{(2)} \\
\end{array}
\right.
\end{equation}
and similarly for $i=k+1,    . . ,    g$. Here $\omega_{p_i}^{i}$
denotes the abelian differential of the second kind with double poles
at $p_i$ \cite{farkas}.
It is then clear that the period matrix behaves as
\begin{equation} \tau\to \left\{
\begin{array}{cc}
\tau^{(1)} & t\gamma \\
 t\gamma & \tau^{(2)} \\
\end{array}
\right \} + O(t^2)
\end{equation}
where $\gamma$ is some constant. The behavior of the prime form
depends on which part of the surface the arguments lie, namely
\begin{eqnarray}
E(z,w) &\to& E^{(i)}(z,    w)+O(t)\mbox{ if }z,w\epsilon\Sigma^{(i)}
\nonumber \\
E(z,w) &\to& t^{-1/2}E^{(1)}(z,p_1)E^{(2)}(w,p_2)+O(t^{1/2})\mbox{ if
}    z\epsilon\Sigma_1,w\epsilon\Sigma_2
\label{eq:degen}
\end{eqnarray}
There is an important difference from the usual open string theory.
Substituting (\ref{eq:degen}) into (\ref{eq:greensf1})  the terms
proportional to $\ln|t|$ cancel between the two contributions because
of the minus sign in (\ref{eq:greensf1}). This means that the
divergent behavior as $t\to 0$ is not momentum dependent and there
are no poles from intermediate open string states. In this respect
the Dirichlet string theory is not a theory of open strings since
there are  only  closed string poles in the scattering amplitudes.
The level $N=0$ and $N=1$ contributions arise from the terms in
$\epsilon_{cl}$ which are of order $t^0$ and $t^1$.
It is easy to see that for the lowest order $\epsilon_{cl}$
factorizes. If the first $l$ vertex operators lie in $\Sigma_1$ and
the rest in $\Sigma_2$ the exponential energy is given by
\begin{equation}
\epsilon_{cl}=\sum_{i,    j=1}^{l}k_ik_jG^{(1)}(z_i,    z_j)+\sum_{i,
   j=l+1}^{n}k_ik_jG^{(2)}(z_i,    z_j) +o(t)
\end{equation}
The possible cross terms vanish to this order since  $I(p_i)=p_i$ and
so the contribution of the first terms in (\ref{eq:greensf2}) cancel.
The factorization of the period matrix leads to the cancellation of
the second terms in (\ref{eq:greensf2}) up to this order.
The $N=0$ contribution can be written
\begin{equation}
\int\frac{dt}{t^2}\langle V(z_1). . V(z_l) \oint\bf{1} \rm
\rangle_{\Sigma^{(1)}}\langle V(z_{l+1}). .
V(z_{n})\oint\bf{1}\rm\rangle_{\Sigma^{(2)}}
\end{equation}
The formulas above can be generalized for the $N=1$ contribution and
there are also formulas for the degeneration of the differentials and
 the prime form under the pinching of nonzero homology cycles
\cite{Fay} which are necessary for a similar treatment of case (c).
One also has to take the factorization of the measure into account
which also gives a nontrivial contribution to the $N=1$ case. In
principle one can do this by using the representation of the measure
(\ref{eq:measure}) in terms of theta functions and prime forms
\cite{verlinde}.  Another  argument is based on the factorization
formulae obtained by Polchinski \cite{pol5}  justifies the formula
(\ref{eq:sewing}).
The behavior of string amplitudes can be investigated by using sewing
techniques in CFT \cite{Sonoda}. For the correlation function on a
closed surface with a fixed conformal structure, a plumbing fixture
construction gives the factorization
\begin{equation}
\langle V_1..V_n\rangle_{\Sigma}=\sum_i
t^{h_i}\overline{t}^{\overline{h_i}}\langle
V_1..V_lA_i\rangle_{\Sigma_1}\langle
V_{l+1}..V_{n}A^i\rangle_{\Sigma_2}
\end{equation}
This amounts to an insertion of a complete set of states at the
punctured surfaces $\Sigma_1$ and $\Sigma_2$ where $h_i$ and
$\overline{h}_i$ are the left and right conformal dimensions of the
state $A_i$.
In string theory one has to integrate over all possible comformal
structures, i.e. over the moduli space. In the formalism with $bc$
ghosts and ghost insertions of (\ref{eq:ghost}) the appropriate thing
to do is to find the Beltrami differential corresponding to the
plumbing fixture coordinates $t,\overline{t}$, pair them with the
ghost zero modes and factorize over the complete CFT including the
ghost states. In the case of a zero (dividing) cycle the other
coordinates on the moduli space are the modular parameters $m^{(i)}$
and the puncture $p^{(i)}$ on the surface $\Sigma_i$,with $i=1,2$. In
the case of a nonzero (nondividing) cycle the moduli parameters are
conveniently chosen to be $m$ of $\Sigma^{\prime}$ and the two
punctures $p^{(i)}$ on $\Sigma^{\prime}$. In order to factorize on a
complete set of states choose a coordinate system $z_i$ at $p^{(i)}$
to get a Hilbert space description of the CFT. The insertion of a
complete set of states in the path integral is given by \cite{pol5}.
\begin{equation}
\sum_it^{h_i-1}\overline{t}^{\overline{h_i}-1}b_0^{(1)}
\overline{b}_0^{(1)}|\Phi_i\rangle^{(1)}|\Phi^i\rangle^{(2)}
\end{equation}
The ghost insertion is $(t\overline{t})^{-1}b_0\overline b_0$
corresponding to the $t,\overline{t}$ coordinates on moduli space.
$|\Phi\rangle^{(i)}$ denotes the insertion of a state $|\Phi\rangle$
in the Hilbert space of states at the puncture $p_i$. The sum is
taken over all states of the combined $X,b,c$ system. There are some
subtleties due to the fact that to define this factorization a choice
of a coordinate system at the punctures has to be made, see
\cite{pol5} for details.
This can be applied to the degeneration of bordered surfaces
discussed in section 2.3,  by noting that closed string factorization
implies the open string result via the doubling procedure. For the
generic example of the half plane geometry the open string conformal
field theory can be expressed by the closed string (bulk) CFT
by identifying $X(\overline z)= \pm \overline{X}(z)$ and just keeping
the chiral half of the theory defined now in the plane. The different
signs correspond to Neumann and Dirichlet boundary conditions.
This local description  works for more complicated topologies since
the sewing is locally described by  coordinate systems at the two
punctures $p_i$.
For the Dirichlet boundary conditions one has to be careful about the
zero modes, the states on which one factorizes are automatically at
zero momentum, as is clear from the discussion at the end of section
2.2.
The factorization on the level one state $N=1$ is then given by (see
appendix B for some conventions for the ghosts)

\begin{equation}
t^{(Y_2-Y_1)^2-1}\mbox{\Large{\{}}-\eta_{\mu\nu}a_{-1}^{(1)\mu}
a_{-1}^{(2)\nu}+c_{-1}^{(1)}b_{-1}^{(2)}-b_{-1}^{(1)}c_{-1}^{(2)}
\mbox{\Large{\}}}c_1^{(1)}|\downarrow\rangle^{(1)}c_1^{(2)}
|\downarrow\rangle^{(2)}
\label{eq:factor}
\end{equation}
The factor of $(Y_1-Y_2)^2$ comes from  $L_0$ and reflects the fact
that the conformal weight of the open string states do not depend on
the momentum but on the boundary positions.
Due to the fact that the expression is at zero momentum  the $a_{-1}$
mode can be separated from the rest which can be written as a BRST
exact state, hence (\ref{eq:factor}) is given by (setting $Y_1=Y_2$)
\[
t^{-1}\mbox{\Large{\{}}-\eta_{\mu\nu}a_{-1}^{(1)\mu}a_{-1}^{(2)\nu}
|\downarrow\rangle^{(1)}c_1^{(2)}|\downarrow\rangle^{(2)}
\]
\begin{equation}
+(Q_{(1)}+Q_{(2)})[b_{-1}^{(2)}|\chi\rangle^{(1)}
|\downarrow\rangle^{(2)}-b_{-1}^{(1)}|\chi\rangle^{(2)}
|\downarrow\rangle^{(1)}]\mbox{\Large{\}}}
\end{equation}
Here the BRST operator on the surface $\Sigma_i$ is denoted $Q^{(i)}$
and
$c_{-1}^{(i)}|\downarrow\rangle^{(i)}=Q^{(i)}|\chi\rangle^{(i)}$,
this is possible because $k^{\mu}=0$ and terms
$a_{-1}^{(i)\mu}k_{\mu}|\chi\rangle^{(i)}$ vanish.
This is remarkable since in the usual case the ghosts are needed to
cancel  unphysical polarizations, here the Dirichlet boundary
condition automatically enforces $k=0$ and the BRST cohomolgy
decouples the ghost states and leaves the level one state. The
appropriate vertex operator insertion corresponding to $a_{-1}^{\mu}$
is given by $\oint\partial_n X^{\mu}$ in the case of Dirichlet
boundary conditions.

The properties of the period matrix under degeneration of a nonzero
cycle can be used to show that this boundary of moduli space is
responsible for the divergences in position space for the Dirichlet
partition function mentioned earlier. If for example $b_g$-cycle is
pinched the period matrix behaves as
\begin{equation} \tau\to \left\{
\begin{array}{cc}
\tau^{\prime} & \oint_{p_1}^{p_2}\omega \\
 (\oint_{p_1}^{p_2}\omega)^t & \ln t \\
\end{array}
\right \} + O(t)
\end{equation}
Here $\tau^{\prime}$ is the period matrix on the genus $g-1$ surface
with the punctures $p_1$ and $p_2$ where the cycle was pinched, the
$\omega_i$ are the remaining $g-1$ holomorphic differentials. Putting
this into (\ref{eq:dxcl}) the $\ln t$ part of the period matrix gives
a term in $\epsilon_{cl}$ which behaves as
$\exp(\ln t (Y_{g-1}-Y_g)^2)$. This term is responsible for the
singularities in $(Y_{g-1}-Y_g)^2$ in the limit $t\to 0$. The above
mentioned singularities outside, on and inside the lightcone are
obtained when the singular behavior of the measure on moduli space is
also taken into account \cite{gava} which contributes $\exp(\ln
t(-1+n))$, where $n$ is a positive integer. The presence of
singularities away from the lightcone is unlike normal field
theoretic amplitudes and this is a decidedly stringy effect
\cite{mbgtsd}.  Other pinches of nonzero cycles  give the
singularities in different combinations of the $Y_i$. In section 4
this is discussed in detail for the case of the annulus. Multiple
pinches should lead to cuts in position space, and this gives  a
similar picture to the analyticity of scattering amplitudes in
momentum space for the Neumann theory, but now in position space.

\section{One Dirichlet boundary insertion}
\label{one}
 The inclusion of one Dirichlet boundary in the world sheet yields
the topology of the half plane. The general method of the last
section is now just the method of images on the half plane, where the
Green function for the complex plane $-1/4\ln|z-w|$ gives for the
Dirichlet Green function
\begin{equation}
G^D(z,    w)=-\frac{1}{4\pi}\{\ln|z-w|^2-\ln|z-\overline{w}|^2\}
\label{eq:greendisk}
\end{equation}
The classical solution is  a constant $X_{cl}^{\mu}=Y^{\mu}$,
which upon integration in (\ref{eq:gauss1}) leads to momentum
conservation in scattering amplitudes.
The scattering amplitude for arbitrary closed string vertex operators
is then given by
\begin{equation}
\Gamma(k_1,    \cdots,
k_n)=\int\frac{\prod_{i=k}^nd^2z_k}{Vol(SL(2,
R))}\langle\mbox{contractions}\rangle \exp\{\epsilon_{cl}\}
\label{eq:disk1}
\end{equation}
Where the exponential energy is given by
\begin{equation}
\epsilon_{cl}=\sum_{i>j}^{n}k_ik_j\{\ln|z_i-z_j|^2-\ln|z_i-\overline
z_j|^2\}+\sum_{i=1}^{n}k_i^2\ln|z_i-\overline z_i|^2
\label{eq:energy}
\end{equation}
the $\langle \mbox{contractions} \rangle$ denote contractions coming
from the tensor
vertex operators.  Vol(Sl(2,R)) denotes the volume of the group
generated by the
conformal Killing vectors,    which can be used to fix $z_1=i$ and to
set $Re(z_2)=0$. The second sum in (\ref{eq:energy}) comes from the
regularization of the Green function at coincident points. The form
of the contractions depends on
the vertex operators considered. The half plane has no modular
parameter so the only divergent contribution can come from case (a)
in the general analysis,    namely, when some of the $z_i$ approach
the boundary, i.e.  $Im(z_i)\to 0$. The analysis of the divergences
in this limit is only a special case of the general analysis of case
(a).

As discussed earlier the high energy fixed angle scattering amplitude
is dominated by the region where $\epsilon_{cl}$ in (\ref{eq:energy})
vanishes. However this coincides with the region where the integral
diverges due to the level zero and and level one open string states.
The naive result is that after fixing $z_1$ with the Moebius group
invariance the integration over deviations $z_2=i\delta y_2,
z_3=x_3+i\delta y_3,    z_4=x_4+i\delta y_4$ gives a power behavior
of $s^{-3}$ since the terms in $\epsilon_{cl}$ are linear in the
deviations. The presence of divergences for external tachyons
invalidates  the analysis.

In order to avoid the problematic divergences we turn to consider
amplitudes with external massless tensor states. Contractions of the
standard massless tensor vertex operators
\begin{equation}
V(k,\xi;z)=:\xi_{\mu\nu}\partial_{z}X^{\mu}(z)
\partial_{\overline{z}}\overline{X^{\nu}(z)}e^{ik_{\rho}X(z,
\overline{z})^{\rho}}:
\end{equation}
give rise to the following terms
\begin{eqnarray*}
\langle\partial_zX^{\mu}(z)\partial_wX^{\nu}(w)
\rangle=\delta^{\mu\nu}\frac{1}{(z-w)^2} \\
\langle\partial_{z}X^{\mu}(z)\partial_{\overline{w}}
\overline{X^{\nu}(w)}\rangle =\delta^{\mu\nu} \frac{1}
{(z-\overline{w})^2}
\end{eqnarray*}
We can check whether there are any
special contractions i.e.  special helicity amplitudes of the tensors
where the infinity above disappears. This is indeed the case for the
following 4-particle scattering amplitude as noted in \cite{mbgtsd}
\begin{equation}
\xi^1_{\mu_1\nu_1}\xi^2_{\mu_2\nu_2}\xi^3_{\mu_3\nu_3}
\xi^4_{\mu_4\nu_4}\langle\partial
X^{\mu_2}\overline{\partial X^{\nu_3}}\rangle \langle \partial
X^{\mu_3}
\overline{\partial X^{\nu4}} \rangle \langle \partial X^{\mu_4}
\overline{\partial X^{\nu_1}} \rangle \langle \partial X^{\mu_1}
\overline{\partial X^{\nu_2}} \rangle
\label{eq:contract}
\end{equation}
To see that such amplitudes are the only ones which are finite define
the following scaling variable $\eta$,    where  either one,    two
or three of $y_2,    z_3,    z_4$ are scaled with $\eta$, defining
$y_2=i\eta$,    $z_3=\eta\xi_3$ and/or $z_4=\eta\xi_4$. If two of the
variables are scaled there is a factor of $\eta^2$ from the measure,
  if three are scaled the contribution is $\eta^4$. It is now easy to
see that there is a total of four contractions for two particle
scattering and that only the `cyclic' type (\ref{eq:contract}) is
finite under these scalings.

There is a general reason that both the $N=0$ scalar and $N=1$ vector
open string state  decouple in an arbitrary diagram contributing to
this particular helicity amplitude. This can be seen by cutting a
diagram where one, two or three vertex operators approach the
boundary. In the case of cyclic contractions there are at least two
Lorentz indices contracted between the two halves of the diagram i.e.
the intermediate propagating state has to have a tensorial structure.
However the open string $N=0$ and $N=1$ states couple to scalars and
derivatives of scalars respectively, and therefore decouple from the
special helicity amplitudes (\ref{eq:contract}). Hence there can not
be a divergence when the vertex operators approach the boundary.

It is easy to see that the considerations are not limited to two
particle scattering and that in  general for $n$ vertex operator
insertions the `cyclic' amplitudes yield no divergence from the
asymptotic region. The high energy fixed angle scattering behavior
for one Dirichlet boundary insertion is governed by the integration
over the deviations of $y_2,    z_3,    . . ,    z_n$ from the real
line which gives $s^{-(n-1)}$ for $n$ vertex operator insertions.


\section{Two Dirichlet boundary insertions}
\label{two}
A world sheet with two Dirichlet boundaries and no handles is the
annulus. The annulus has one real modular parameter (the ratio
between the radii). The torus with purely imaginary modular parameter
$\tau=il$ is represented  by a fundamental cell in the z-plane
\begin{equation}
0<Im(z)<l:0<Re(z)<1
\label{eq:cylind}
\end{equation}
The holomorphic differential is $\omega=dz$ and the period matrix is
given by the modular parameter $\tau$. The annulus can be represented
as  the quotient under the Involution $I(z)=\overline{z}$,    and the
boundaries are given by $\partial\Sigma_1=\{Im(z)=0\}$ and
$\partial\Sigma_2=\{Im(z)=1/2\}$. (See Figure 5)
The measure on the moduli space of the torus is  \cite{dhoker}
\begin{equation}
\int\frac{d^2\tau}{|Im(\tau)|^{14}}|\eta(\tau)|^{-48}
\end{equation}
The presence of conformal killing vectors on the torus and annulus
has to be included in the calculation of the measure and the measure
on the annulus is not given by  (\ref{eq:measure2}),    but
\cite{burgess}
\begin{equation}
\int \frac{dl}{l^{13}} n(il)^{-24}
\label{eq:annmeasure}
\end{equation}
Where the $\eta$-function is given by $\eta(il)=e^{-\pi
l/12}\prod(1-e^{-\pi  nl})$. The Green function on the torus is well
known to be
\begin{equation}
G_{\overline\Sigma}=-\frac{1}{4\pi}\ln\left|\frac{\theta_1(z-w|\tau)}
{{\theta_1}\prime(0|\tau)}\right|^2+\frac{1}{2}\frac{\{Im(z-w)\}^2}
{Im\tau}
\end{equation}
the second term comes from
$\frac{1}{2}Im\int^w_z dz Im\tau ^{-1}\int^w_z
dz=\{Im(z-w)\}^2/Im\tau$. the
Dirichlet Green function on $\Sigma$ can then be expressed
using (\ref{eq:modgreen1}) as
\begin{equation}
G^D(z,w)=\frac{1}{4\pi}\ln\left|\frac{\theta_1(z-w|\tau)}
{\theta_1(z-\overline{w}|\tau)}\right|^2-2\frac{(z-\overline{z})
(w-\overline{w})}{Im\tau}
\label{eq:torusd}
\end{equation}

In this case $X_{cl}(z)=Y_1-(Y_2-Y_1)/Im(\tau)$ is the classical part
in (\ref{eq:classical1}). The integral  over $Y_1$ and $Y_2$ is done
as in (\ref{eq:dirich1}) using the relations
 \begin{eqnarray}
 b & = & \sum_i k_i\int_0^1\partial_w G^D(z_i,    w) \nonumber \\
    & = & \sum_i k_i\frac{(z_i-\overline{z}_i)}{Im\tau}
\label{eq:torusd1}
\end{eqnarray}
The second term in (\ref{eq:torusd}) is cancelled by the contribution
coming from (\ref{eq:torusd1}) inserted in (\ref{eq:modgreen1}) and
the modified Green function is given by
\begin{equation}
\hat G^D(z,w)=\frac{1}{4\pi}\ln\left|\frac{\theta_1(z-w|\tau)}
{\theta_1(z-\overline{w}|\tau)}\right|^2
\label{eq:angreen}
\end{equation}
 which coincides with the Green function found in \cite{li} obtained
with different methods. The integration produces  powers of
$Im(\tau)$ so that the measure (\ref{eq:annmeasure}) is given by
\begin{equation}
\int dl n(il)^{-24}
\label{eq:intmeasure}
\end{equation}
\subsection{High-energy fixed angle scattering of massless tensor
states}
High energy fixed angle scattering is dominated by the contributions
of the endpoints at which the vertices are close to  the boundary.
The behavior of the Green function $\hat G^D(z,    w)$ is given by
(we denote the modified Green function of section 4  by G)
\begin{eqnarray}
z\epsilon \partial\Sigma_1\mbox{ , }    w\epsilon \partial\Sigma_1
&:&
G^D=0\mbox{ , }    \partial_z G^D=0\mbox{ , }    \partial_z
\partial_w G^D \neq 0 \nonumber \\
z\epsilon \partial\Sigma_1\mbox{ , }    w\epsilon\partial\Sigma_2 &:&
G^D
=0\mbox{ , }    \partial_z G^D=\frac{i\tau}{2}\mbox{ , }
\partial_z\partial_w G^D \neq 0 \nonumber \\
z\epsilon \partial\Sigma_2\mbox{ , }    w\epsilon\partial\Sigma_2 &:&
G^D=\frac{1}{2}\mbox{ , }    \partial_z G^D=\frac{i\tau}{2}\mbox{ , }
   \partial_z \partial_w
G^D \neq 0
\label{eq:boundgreen}
\end{eqnarray}
Note that there seems to be an asymmetry between the two boundaries
in the formula above, this is not so in the scattering amplitudes
because of momentum conservation.
As in the case of the one boundary insertion  the scattering
amplitudes are constructed from the
Green function the `exponential energy' is given by
\begin{equation}
\epsilon_{cl} = \pi\alpha^{\prime}\sum_{i\neq j}G^D(z_i,
z_j)k_ik_j
\end{equation}
In order to analyze the behavior of $\epsilon$ when all the $z_i$ are
near
the boundary $\partial\Sigma_1$.
 write $z_i=x_i + i\delta y_i$ and make a Taylor expansion at $\delta
y_i=0$.
Using the fact that on the boundary $z_i=\overline z_i$ it is easy to
see that the only term up to order $o(\delta y^2)$ is
\begin{equation}
G^D(z_i,    z_j)=\frac{1}{2}\partial_{y_i}\partial_{y_j}G^D(x_i,
x_j)\delta
y_i\delta y_j + o(\delta y^2)
\end{equation}
Using complex coordinates $\frac{\partial}{\partial
y}=i\frac{\partial}{\partial z} - i\frac{\partial}{\partial\overline
z}$ and defining
\begin{equation}
h(z,w)=-\frac{1}{4\pi}\frac{\theta_1^{\prime\prime}(z-w|\tau)
\theta_1(z-w|\tau)-\theta_1^{\prime}(z-w|\tau)^2}
{\theta_1(z-w|\tau)^2}
\label{eq:hzw}
\end{equation}
Defining $\partial_{y_i}\partial_{y_j}G(x_i,    x_j)=H_{ij}$ we see
that
\begin{eqnarray}
H_{ij}  & = & -4h(x_i,    x_j) \label{line1} \\
 & = & -\pi\{\frac{1}{\sin ^2\pi (x_i-x_j)}+ 8\sum_n
n\frac{q^{2n}}{1-q^{2n}}\cos 2n\pi(x_i-x_j)\} \label{line2}
\end{eqnarray}
where  the following  identity \cite{bateman} for theta functions was
used
\begin{equation}
\frac{\theta_1^{\prime}(z|\tau)}{\theta_1(z|\tau)} = \pi\cot{ \pi z}
+
4\pi\sum_n\frac{q^{2n}}{1-q^{2n}}\sin {2m\pi z}
\end{equation}
It is now simple to write $\epsilon_{cl}$ as a power series in
$\delta y_i$
for $i=1,    \cdots,    n$.  In the fixed angle high energy limit
which we want
to consider,     all factors $k_ik_j$ become large. A end point
integration of  the deviations $\delta y_i$ gives the asymptotic
behavior
of the scattering amplitude. For the two particle scattering
with the Mandelstam variables $s=-(k_1+k_2)^2,    t=-(k_1+k_3)^2,
u=-(k_1+k_4)^2$
\begin{eqnarray}
\epsilon_{cl} & = &
s\{H_{12}\delta y_1 \delta y_2 +H_{34}\delta y_3 \delta y_4 \}+
t\{H_{13}\delta y_1\delta y_3+H_{24}\delta y_2 \delta y_4\}
\nonumber \\
 & & +u\{H_{14}\delta y_1 \delta y_4 +H_{23}\delta y_2\delta y_3\}
 \nonumber \\
 & = & \sum_{ij}M_{ij}\delta y_i\delta y_j
\end{eqnarray}
The Gaussian integration gives a factor $(\det M)^{-1/2}$. Evaluating
the determinant and noting that for $s\rightarrow\infty$  in the
fixed angle limit
$-s/t=\sin^2\Phi/2$ and  $-u/s=\cos^2\Phi /2$ where $\Phi$ is the
centre of mass frame scattering angle, the determinant is given by
\begin{eqnarray}
\det(M) & = &
s^4\{(H_{12}H_{34})^2+(H_{13}H_{24})^2\sin^8\frac{\Phi}{2}
+(H_{14}H_{23})^2\cos^8\frac{\Phi}{2} \nonumber \\
 & &+
2H_{12}H_{34}H_{23}H_{14}+2H_{12}H_{34}H_{13}H_{24}
\cos^4\frac{\Phi}{2} \\  & &
-2H_{23}H_{13}H_{14}H_{24}\cos^4\frac{\Phi}{2}\sin^4
\frac{\Phi}{2}\} \nonumber
\end{eqnarray}
The term in the brackets  depends  only on the scattering angle.
Therefore
we expect that the leading asymptotic behavior of the scattering
amplitude is $s^{-2}$. The explanation for the smaller power in
comparison to the disk topology lies in the fact that for the disk
there is the $SL(2,    R)$ conformal Killing group,    which allowed
to fix one of
the vertices at the origin of the disk,    so there could only be
three
vertices touching the boundary.
 If one vertex touches one boundary and three are close to the other
$\epsilon_{cl}$ is linear in the deviations and the asymptotic
behavior is  $s^{-3}$ which is subleading. If two vertices  are on
one boundary and two are on the other $\epsilon_{cl}$ does not vanish
and the amplitude is exponentially suppresed.

\subsection{Boundary of moduli space}
 This  analysis is naively correct but one has to consider the effect
of possible divergences coming from the boundary of moduli space. The
cases to consider are  (a): vertices going to $\partial\Sigma_1$ or
$\partial\Sigma_2$,    case(c): the limit $l\to0$ and  case (d): the
limit $l\to\infty$. \\[.25  in]
(a) As for the disk the tachyon amplitudes are meaningless due to a
boundary divergence since the self-contraction $:G^D(z,z):$ is
defined
by
\begin{equation}
\lim _{z\rightarrow w}\left\{
G^D(z,w)-\frac{1}{4\pi}\ln|z-w|^2\right\}=\frac{1}{4\pi}\ln
\left|\frac{\theta_1^{\prime}(0|\tau)}{\theta_1(z-\overline
z|\tau)}\right|^2
\end{equation}
and this behaves like $|z-\overline z|^2$ if z is near the
boundary. As we shall see the `cyclic' tensor-amplitudes like
(\ref{eq:contract}) circumvent these divergences.

The contractions for the tensor-amplitudes are given by
(\ref{eq:hzw})
\begin{equation}
\langle\partial_zX\partial_wX\rangle=h(z, w)
\end{equation}
\begin{equation}
\langle\partial_zX\overline{\partial_wX}\rangle = h(z,\overline
w)
\end{equation}
To consider the finiteness of this amplitude we can repeat the
analysis of section \ref{one} for case (a) which applies also for the
annulus
(since near a particular boundary the other boundaries can be
ignored). But  now all
four vertices  can come close to one boundary. This region of moduli
space
can be investigated defining new variables
\begin{equation}
z_1=x_1+i\eta,z_2=x_1+\eta\xi_2,z_3=x_1+\eta\xi_3,
z_4=x_1+\eta\xi_4
\end{equation}
 The measure
behaves as $\prod_{i=1}^{3}d^2z_i=\eta^6d\eta dx_1 \prod_{i=2}^{4}d^2
\xi_i$ in terms of the new integration variables. The Green function
behaves as
\begin{equation}
G^D(z_i,z_j)=-1/4\pi\ln\left|\frac{\xi_i-\xi_j}
{\xi_i-\overline{\xi_i}}\right|^2+o(\eta^2)
\end{equation}
(note that $\xi_1 = i$ and that there is no linear term in $\eta$).
In contrast to the analysis of the half plane  the four vertices give
a
contribution which is singular,    since there is a $\eta^{-8}$
coming from the contractions of (\ref{eq:contract}) $\int d\eta
\eta^{-2}A_1$. The divergent part
 $A_1$ is given by
\begin{eqnarray}
A_1 & = & \int dl\mu(l)
\int\prod_{i=1}^{4}d^2\xi_i\delta^2(\xi_1-i)\frac{1}
{(\xi_1-\overline{\xi}_2)^2}\frac{1}{(\xi_2-\overline{\xi}_3)2}
\nonumber
\\
& &
\frac{1}{(\xi_3-\overline{\xi}_4)^2}\frac{1}
{(\xi_4-\overline{\xi}_1)^2}\exp\{\sum_{i\neq
j}k_ik_jln\left|\frac{\xi_i-\xi_j}{\xi_i-\overline{\xi}_j}
\right|^2\}
\end{eqnarray}
$\mu(l)$ is given by (\ref{eq:intmeasure}).  Note that $A_1$ is very
similar to the amplitude for one boundary insertion. This means that
the high energy fixed angle
behavior in the limit in which all four vertices come together is
divergent as $\int d \eta \eta^{-2}$.  The coefficient multiplying
this divergence has a $s^{-3}$ dependence,    hence it is
not a leading contribution. A sensible regularization should not
spoil
this feature.
This divergence comes from the presence of the open string $N=0$
state as
explained above. There is no divergence coming from the $N=1$ state
since the next order term in expanding the Green function
is quadratic in $\eta$. This is easy to explain since  the $N=1$
state couples to the momenta flowing through the boundary,    when
four vertices come  together the degeneration in case (a) leads to a
disk with four vertices coupling to the rest of the diagram with no
vertices at all,  so that zero momentum flows through the boundary
connecting the two parts of the diagram because of momentum
conservation.\\[.15  in]
(c) The limit $l\to 0$ corresponds to the limit in which the  radii
of the annulus coincide. To investigate this boundary of moduli space
it is useful to make a modular transformation with
$il^{\prime}=\tau^{\prime}$
\begin{equation}
\tau^{\prime}=-\frac{1}{\tau} \mbox{ };\mbox{ }
\zeta_{z}=-\frac{z}{\tau}
\end{equation}
The fundamental cell for the annulus is mapped into $0<Re(\zeta)<1/2$
,     $0<Im(\zeta)<l^{\prime}$ (see Figure 5). The transformed terms
can be expressed using the well known properties of the
$\eta$-function and $\theta$-function under modular transformations.
The resulting  measure is
\begin{equation}
\int dl\eta(il)^{-24} = \int \frac{ dl^{\prime}}{l^{\prime
14}}\eta(il^{\prime})^{-24}
\end{equation}
It can be seen that this limit is also responsible for the
singularities in the Dirichlet partition function $A(Y_1-Y_2)$ for
two boundaries. In this case the general formula (\ref{eq:partition})
reads in the new variables
\begin{equation}
A(Y_1-Y_2)=\int_0^{\infty}\frac{dl^{\prime}}{l^{\prime}}
\eta(il^{\prime})^{-24}\exp\{-\frac{1}{2\pi\alpha^{\prime}}l^{\prime}
(Y_1-Y_2)^2\}
\label{eq:anpart2}
\end{equation}
In the limit $l^{\prime}\to\infty$ expanding the $\eta$-function in
powers of $\exp(2\pi l^{\prime})$ and get $\eta(il^{\prime})^{-24}\to
\exp(2\pi l^{\prime})(1+o[\exp(-2\pi l^{\prime})]$. There are
logarithmic divergences for certain values of $(Y_2-Y_1)^2$ namely
for $(Y_2-Y_1)^2=4\pi^2\alpha^{\prime}(1-N)$ where $N$ is an non
negative integer. The $N=0$ term is a position space singularity
outside the lightcone and the $N=1$ is on the lightcone. The terms
with $N>1$ give an infinite tower of singularities inside the light
cone.
The Green function is written in terms of the new variables using the
well
known imaginary transformation of Jacobi \cite{bateman}
\begin{equation}
\theta_1(-\frac{z}{\tau}|-\frac{1}{\tau})=-i(-i\tau)^{\frac{1}{2}}
\exp(\frac{i\pi
z^2}{\tau})\theta_1(z|\tau)
\end{equation}
The Green function (\ref{eq:angreen}) written in the new arguments is
then easily calculated,
\begin{equation}
G^D(\zeta_z,
\zeta_w)=-\frac{2i}{\tau^{\prime}}Re(\zeta_z)Re(\zeta_w)
-\frac{1}{4\pi}\ln\left|\frac{\theta_1(\zeta_z-\zeta_w|
\tau^{\prime})}{\theta_1(\zeta_z+\overline{\zeta}_w|\tau^{\prime})}
\right|^2
\label{eq:bound1}
\end{equation}
Note the change of sign in the theta function in the denominator
which
is due to the fact that $\overline{\tau}=-\tau$.
To discuss the high energy fixed angle scattering  it is convenient
to make yet another change of variables. First map  the fundamental
cell onto the seminannular region with an exponential map
$\sigma_i=\exp(2\pi\zeta_i)$ and  then introduce  new variables $x_i$
instead of the $\sigma_i,    i=1,    . . ,    n$ (see figure 5)
\begin{eqnarray}
\sigma_1 &=& x_1 \nonumber \\
\sigma_2 &=& x_1 x_2\nonumber \\
&:& \nonumber \\
\sigma_n &=& x_1. . x_n
\label{eq:sigmas}
\end{eqnarray}
The rotational conformal Killing symmetry of the integrand ( in the
$\zeta$ plane this amounts to  the invariance of the integrand under
imaginary translations)  can be used to fix $\sigma_n=|w|$. With the
exponentiated modular parameter $|w|=\exp(-2\pi l^{\prime})$. The
region $\{x_i:|x_i|<1,    Im(x_i)>0,    i=1,    . . n\}$ covers the
moduli space of the semiannular parameterization where the $\sigma_i$
are ordered,   i.e.$|\sigma_1|>|\sigma_2|>. . >|\sigma_n|$. To cover
the whole moduli space all  orderings of the $\sigma_i$ in
(\ref{eq:sigmas}) have to be considered.
The Jacobian for this change of variables gives
\begin{equation}
\int \frac{dl}{{l}^{14}} \eta(il)^{-24}\prod_{i=1}^n
d^2\zeta_i=\prod_{i=1}^{n}d^2
x_i|w|^{-3}\prod_{k=1}^{\infty}(1-|w|^k)^{-24}\ln|w|^{-14}
\label{eq:wmeasure}
\end{equation}
The case (c) corresponds to $|w| \to 0$. The new variables are useful
since one can distinguish between different limits depending on which
and how many of the $x_i$ go to zero. The $x_i$ are the string
equivalent of Schwinger parameters in ordinary field theory. The
limit in which one of them vanishes  corresponds to a pinching of an
internal propagator (\ref{eq:propagator}). This becomes clear if  the
annulus amplitude is constructed in an operator approach as a trace
of vertex operators and propagators  $Tr(V\Delta. . V\Delta)$ and the
parameterization (\ref{eq:sigmas}) is used \cite{gsw}.

The arguments which appear in the Green function can be written in
terms of the $\xi_i$ for $i>j$.
\begin{eqnarray}
\xi_i &=& \frac{1}{2\pi i} \ln (x_1. . x_i) \nonumber \\
\xi_i-\xi_j &=& \frac{1}{2\pi i} \ln(\frac{x_1. . x_i}{x_1. . x_j})
\nonumber \\
\xi_i+\overline{\xi}_j &=& \frac{1}{2\pi i} \ln(\frac{x_1. .
x_i}{\overline{x}_1. . \overline{x}_j})
\end{eqnarray}
In the limit $w\to 0$  some of the $|x_i|\to 0$. The first term of
the Green function (\ref{eq:bound1}) can be dominated by a finite,
 $w$ independent contribution from the second term. To see this
consider the expansion of $\theta_1$ \cite{bateman}
\begin{equation}
\theta_1(\xi|\tau)=C\sin\pi \xi
\prod_{n=1}^{\infty}(1-2w^{n}\cos2\pi \xi +w^{2n})
\label{eq:theta1}
\end{equation}
Here $C$ is a $w$ dependent term which cancels between the two
$\theta$-functions in the Green function. In the limit under
consideration only $\sin \xi$ and the $n=1$ factor in
(\ref{eq:theta1}) contribute. They appear in the expression for the
Green function in the following ratios
\begin{eqnarray}
\left|\frac{\sin\pi (\xi_i-\xi_j)}{\sin\pi
(\xi_i+\overline{\xi}_j)}\right|^2 &=& \left| \frac{1-\frac{x_1. .
x_i}{x_1. . x_j}}{1-\frac{x_1. . x_i}{\overline{x_1}. .
\overline{x_j}}}\right|^2 \mbox{ if } \frac{x_1. . x_i}{x_1. .
x_j}\neq 0 \\
\left|\frac{1-w\cos2\pi (\xi_i-\xi_j)}{1-w\cos2\pi
(\xi_i+\overline{\xi}_j)}\right|^2 &=& \left| \frac{1-w\frac{x_1. .
x_j}{x_1. . x_i}}{1-w\frac{x_1. . x_j}{\overline{x_1}. .
\overline{x_i}}}\right|^2 \mbox{ if } w\frac{x_1. . x_i}{x_1. .
x_j}\neq 0
\label{eq:bound2}
\end{eqnarray}
In all other cases the ration of $\theta$ functions is one in this
limit and the second term in (\ref{eq:bound1}) vanishes. Writing
$\epsilon_{cl}=\epsilon_{-1}+\epsilon_{0}$ up to terms vanishing in
linear order of $|x_i|$. Firstly $\epsilon_{-1}$ is coming from the
first term of (\ref{eq:bound1}).
\begin{equation}
\epsilon_{-1}=\sum_{ij}k_ik_j\frac{1}{2\pi^2\ln w}\ln(\frac{x_1. .
x_i}{\overline{x}_1. . \overline{x}_i})\ln(\frac{x_1. .
x_j}{\overline{x}_1. . \overline{x}_j})
\label{eq:eps1}
\end{equation}
Secondly $\epsilon_{0}$ is coming from the nonvanishing terms in
(\ref{eq:bound2}). Consider the example in which $x_1\to0$ and all
other $x_i$ finite. Writing $x_1=\eta e^{i\phi_1}$ with $\eta\to 0$
\begin{eqnarray}
\epsilon_0 & = &
k_1k_2\ln\left|\frac{1-x_2}{1-x_1x_2/\overline{x}_1}\frac{1-x_4}
{1-(x_1x_2x_3x_4)/\overline{(x_1x_2x_3)}}\right|^2 \nonumber \\
 & + & k_1k_3\ln\left|
\frac{1-x_2x_3}{1-x_1x_2x_3/\overline{x_1}}\frac{1-x_3x_4}
{1-(x_1x_2x_3x_4)/\overline{(x_1x_2)}}\right|^2 \nonumber \\
 & + &
k_1k_4\ln\left|\frac{1-x_2x_3x_4}{1-x_1x_2x_3x_4/\overline{x}_1}
\frac{1-x_3}{1-(x_1x_2x_3)/\overline{(x_1x_2)}}\right|^2\nonumber \\
& + & o(\eta)
\label{eq:epsil0}
\end{eqnarray}
The first three terms in (\ref{eq:epsil0}) depend only on
$x_1/\overline{x_1}$ and are independent of $\eta$. Denoting $x_i=r_i
e^{i\phi_1}$ for $i=2,3,4$ it can be easily seen that the $\eta$
independent part of $\epsilon_0$ has still the property we met in the
general analysis of section 4.1 namely that it vanishes when the
vertex operator positions approach the boundaries. The boundary
$\partial\Sigma_1$ corresponds to $\phi_i=0$ and $\partial\Sigma_2$
corresponds to $\phi_i=\pi$ for all $i$ (see figure 5).
It can also be shown that the $\eta$-independent part of $\epsilon_0$
is quadratic in the deviations $\delta\phi_i$ at the boundaries.
The measure (\ref{eq:wmeasure}) gives a contribution $\int_0d\eta
\eta^{-2}(\ln\eta)^{-14}$ which is divergent at the lower integration
limit and is caused by the $N=0$ state of our general analysis.
The $N=1$ state is given by  the next order in $\eta$. This gives a
$\int_0d\eta\eta^{-1}(\ln\eta)^{-14}$ which is finite at the lower
limit. Note that for the finiteness it was crucial to integrate over
the $Y_1-Y_2$ since this produces the necessary powers of $\ln\eta$.
Without the integration the contribution would be $\int d\eta
\eta^{-1} (\ln\eta)^{-1}$, which diverges.
The term $\epsilon_{-1}$ given by (\ref{eq:eps1}) is proportional to
$1/\ln\eta$
and vanishes in the limit $\eta\to 0$, it can therefor be disregarded
in comparison to $\epsilon_0$.

For  high energy fixed angle scattering the following situation
emerges: The power behavior in $s$ is determined by the
$\eta$-independent part of $\epsilon_0$  and gives the generic
behavior $s^{-2}$ which was found away from the boundary of moduli
space. The divergence due to the $N=0$ state does not spoil this
feature.
This analysis does not depend on the specific example chosen as long
as one of the $x_i$ is finite.\\[.25  in]
(d) This limit corresponds to the inner radius of the annulus going
to zero in the parameterization of (\ref{eq:cylind}) this is given by
 $l\to\infty$. One can view this as a closed string tadpole were a
closed string state propagates along a cylinder of length $l$ and
couples to a disk. The divergences come from the closed string
tachyon and dilaton at zero momentum propagating for an infinite long
time. These are the usual divergences from the closed string sector
which are not special to Dirichlet strings.  Writing $q=e^{-\pi l}$
the measure (\ref{eq:intmeasure}) gives $\int dq
|q|^{-3}\prod(1-|q|^{2n})^{-24}$ and there are divergent terms coming
from $|q|^{-3}$ (tachyon) and $|q|^{-1}$ (dilaton).
To consider the limit $l\rightarrow \infty$  the well known
representation of the theta function \cite{bateman} can be used
\begin{equation}
\theta_1(z|\tau)=2q^\frac{1}{2}\sum_{n=0}^{\infty}(-1)^nq^{n(n+1)}
\sin(2n+1)\pi z
\end{equation}
Where $q$ is given by $q=e^{i\pi\tau}$. To lowest order in q  the
q independent term for the Green function $G^D(z,    w)$ is given by
(\ref{eq:angreen})
\begin{equation}
G^D(z,    w) \longrightarrow\frac{1}{4\pi}
\ln\left|\frac{\sin\pi(z-w)}{\sin\pi (z-\overline w)}\right|^2+o(q^2)
\end{equation}
Mapping the z-plane into the $\rho$-plane via $\rho=\exp(2\pi i z)$
 the cylinder is mapped into the annulus (see figure 5), in this
conformal frame the
Green function has the form
\begin{equation}
G^D(\rho_z,
\rho_w)\longrightarrow\frac{1}{4\pi}\ln\left|\frac{\rho_z-\rho_w}
{1-\rho_z\overline\rho_w}\right|^2+o(q^2)
\label{eq:cased}
\end{equation}

To analyze the fixed angle high energy scattering power behavior in
this limit, note that the Green function  has a $q$ independent part
which in the parameterization of (\ref{eq:cased}) has the form of the
Green function for the half plane (\ref{eq:greendisk}) transformed
into the conformal frame of the disk. Hence the properties of the
Green function discussed in section 3 are valid. The closed string
tachyon and dilaton give divergent contributions, but the fixed angle
high energy behavior is governed by the $q$-independent part of
(\ref{eq:cased}) and a similar analysis to section 3 and 4 shows that
the behavior is still $s^{-2}$ (no vertex operator position has been
fixed in contrast to section 3). The regularization of the divergence
does not spoil this feature.\\[.15in]
To summarize we stress that the boundary of moduli space potentially
invalidates the analysis of the power behavior of the high energy
fixed angle scattering because of the divergences which occur here.
For the annulus it has been shown that the divergent contribution and
the part responsible for the power behavior can be disentangled and
the analysis at generic points of the moduli space is still valid in
this limit.
\section{Conclusions}

In this paper we have analyzed the inclusion of Dirichlet boundaries
perturbatively in a path integral framework. The inclusion of a
finite number of Dirichlet boundaries gives a string theory with
novel features and the following points seem to be generic for an any
number of boundaries.

High energy fixed angle scattering amplitudes decrease with a power
of the center of mass energy. Our analysis of the annulus should
carry over to an arbitrary number of boundaries. Generically the
Green function vanishes when all vertices are on one boundary and is
quadratic in the deviations. The generic behavior for n vertex
insertions is then given by $s^{-n/2}$ from the Gaussian integral
over the deviations.
In contrast to the analysis of high energy fixed angle scattering for
conventional  closed strings \cite{gross1} this behavior comes from a
boundary effect and there is no dominating saddle point. This means
that there is no simple  classical trajectory which dominates the
functional integral in a semiclassical approximation.   The
divergences and the integration over moduli space have to be
considered carefully, which makes it difficult to give universal
statements for all orders.

 By considering special helicity amplitudes for the scattering of
massless tensor states the $N=0$ and $N=1$ divergences have been
avoided in the discussion of the fixed angle high energy scattering.
The divergences which are analyzed in section \ref{divergences} are a
novel feature of Dirichlet strings and there are different ways to
make sense out of them. The general r\^ole of these divergences
should be studied further. One way to get rid of the $N=0$ divergence
might be to consider a supersymmetric version of a theory with
Dirichlet boundary insertions \cite{mbg8}.

In general we have to deal with the $N=1$  divergence in the theory.
There are two different suggestions of how to deal with them. In one
of them  \cite{mbg7},    the $N=1$ state is interpreted as a Lagrange
multiplier  field which has to be integrated over (see appendix B),
in the other \cite{pol3},  a Fischler-Susskind type of mechanism
arises, where the divergences cancel between diagrams of different
topology (see appendix C for a brief description in our context).

Our investigation was limited to a perturbative treatment with a
fixed number of boundary insertions. The natural and very difficult
question is what happens if one sums over arbitrary many boundary
insertions,   i.e. generating  a condensate of Dirichlet boundaries.
It might be possible to make progress by looking  at subcritical
string theory or matrix models,    where one has a handle on
nonperturbative questions and one might be able to take the
condensation of boundaries into account. Another advantage is that
2-d string theory is a consistent bosonic theory so the divergences
coming from the tachyon which are there in critical bosonic string
theory are avoided.

Since one motivation for the investigation of Dirichlet string theory
is the search for a string theory of QCD,    these nonperturbative
questions are very important because perturbatively  the massless
graviton is till present in the spectrum.
\\[.25in]
\bf{Acknowledgements}\\
\rm
I would like to thank  M.B. Green for a lot of useful discussions and
explanations. I would also like to thank the Theory-Division at CERN
for hospitality. This work was supported in part by EPSRC and a
Parnett Research Studentship of Churchill College, Cambridge.
\appendix{\label{modapp}
\section{Modification of the Green function}
This appendix will consider  the term $f$  which modifies the Green
function
$\hat G^D=G^D+f$, from (\ref{eq:modgreen1})
\begin{equation}
f(z_i,    z_j)=2\pi\alpha^{\prime}\sum_{k,
l}\oint_{a_k}\partial_{w_1}G^D(z_i,
w_1)Im\tau_{kl}\oint_{a_l}\partial_{w_2}G^D(z_j,    w_2)
\label{eq:fzw}
\end{equation}
With the Green function given by (\ref{eq:greensf1}), $\partial_z
G^D(z,    w)$ can be expressed in the following way
\begin{equation}
\partial_z G^D(z,    w)=-\frac{1}{4}\partial_z \ln
E(z,    w)-\frac{i}{2}\sum_{i,
j}\omega_i(z)Im\tau_{ij}^{-1}\{Im\int_z^w\omega_j-Im\int_z^{I(w)}
\omega_j\}
\label{eq:harasam}
\end{equation}
Note that the line integral can be rewritten into $\int_w^{I(w)}$
which is independent of z. The derivative of the prime form E gives a
holomorphic differential of the second kind which  vanishes when
integrated along a-cycles,   using the normalization (\ref{eq:diff})
the line integral around an $a_l$-cycle of (\ref{eq:harasam}) is
given by
\begin{equation}
\int_{a_l}dz\partial_zG(z,    w)=\frac{i}{2}\sum_jIm\tau_{lj}^{-1}Im
\int_w^{I(w)}\omega_j
\end{equation}
Using this to  evaluate (\ref{eq:fzw}) gives
\begin{equation}
f(z_i,    z_j)=-\frac{1}{4}\sum_{kl}\int_{z_i}^{I(z_i)}\omega_k
Im\tau_{kl}^{-1}\int_{z_j}^{I(z_j)}\omega_l
\end{equation}
We can now reexpress this via the identity
\begin{equation}
-4\int_{z}^{I(z)}\omega
\int_w^{I(w)}\omega=\left(\int_z^{I(z)}\omega-\int_w^{I(w)}
\omega\right)^2-\left(
\int_z^{I(z)}\omega+\int_w^{I(w)}\omega\right)^2
\end{equation}
where the indices of the $\omega_i$ are omitted which cannot cause
any
confusion since  the matrix $Im\tau^{-1}$ is symmetric.
Now rewrite the line integrals in the following fashion
\begin{eqnarray}
\left (\int_z^{I(z)}-\int_w^{I(w)}\right)\omega & = & \left(
\int_z^{I(z)}+\int_{I(z)}^{w}-\int{I(z)}^{I(w)}\right)\omega
\nonumber \\
 & = & 2i Im\int_z^w\omega \nonumber \\
\left (\int_z^{I(z)}+\int_w^{I(w)}\right)\omega & = & \left(
\int_z^{I(z)}+\int_{I(z)}^{I(w)}-\int{I(z)}^{w}\right)\omega
\nonumber \\
 & = & 2i Im\int_z^{I(w)}\omega
\end{eqnarray}
Note that these quantities are independent of the homology, i.e. they
are single valued functions on the Riemann surface because the
dependence on the homology cancels between the contributions.
With the identity above we see that $f(z_i,    z_j)$ can be cast in
the form
\[
f(z_i,z_j)=\frac{\pi\alpha^{\prime}}{2}\sum_{kl}Im\int_{z_i}^{z_j}
\omega_{k}Im\tau_{kl}^{-1}Im\int_{z_i}^{z_j}
\]
\begin{equation}
-\sum_{kl}Im\int_{z_i}^{I(z_j)}\omega_{k}Im\tau_{kl}^{-1}
Im\int_{z_i}^{I(z_j)}
\end{equation}
This is exactly the quantity needed for the modification in
(\ref{eq:modgreen1}).

\section{The r\^ole of the $N=1$ state}

This appendix will review the r\^ole of the level one state in
Dirichlet string theory. The  factorization on a dividing strip (case
b of section 2.3) does not give open string poles in the intermediate
states as in the theory with Neumann boundary conditions. It has been
suggested in \cite{mbg7}  that the $N=1$ state is really a Lagrange
multiplier field    and that the divergences are caused by the fact
that there is no kinetic  term for Lagrange multiplier fields and its
propagator becomes singular. The issue concerns the quantization of a
free open string on an infinite strip (as in (\ref{eq:dirquant}))
with boundary condition $Y_1=Y_2$ - both boundaries are at the same
point in space time. The zero mode of the energy momentum tensor is
then independent from the position of the boundary $Y$ : $L_0=N$.
The physical state condition on a state $|\Phi\rangle$ is
$(L_0-1)|\Phi\rangle =0 $ and is trivially satisfied for
$|\Phi\rangle=\omega(Y)_{\mu}a_{-1}^{\mu}|o\rangle$. So there is no
condition on the wave function $\omega(y)_{\mu}$. This becomes
clearer if  the reparameterization ghosts are included and the
BRST-cohomology of free string fields is used.
 The BRST-operator \cite{gsw} for Dirichlet strings is given by
\begin{equation}
Q_{BRST}=:\sum_m c_m\{L^{X}_{-m}+\frac{1}{2}L^{gh}_{-m}-1\delta_{m,
 0}\}
\end{equation}
Where $L^X_m$ is given by the modes of the $X$-part of the energy
momentum tensor and $L^{gh}_m$ are the usual
modes of the bc-ghosts energy-momentum tensor. Since the boundary
conditions of the space time fields do not influence the local
geometry on the world sheet, the bc-ghost do not feel the Dirichlet
boundary conditions. The most general field of ghost number
$-\frac{1}{2}$ at level 1 is given by the following field
\begin{equation}
(\omega(Y)_{\mu}a_{-1}^{\mu}+\lambda(Y)c_0b_{-1})|\downarrow\rangle
\end{equation}
(following the conventions of \cite{gsw}). The state
$|\downarrow\rangle$
is the state annihilated by all positive ghost modes and has ghost
number $-\frac{1}{2}$. For free string-fields$|\Phi\rangle$ the
condition
$Q_{BRST}|\Phi\rangle =0$ generates the linearized equations of
motion
for the wavefunctions $\omega(Y)$ and $\lambda(Y)$. Gauge
transformations on the Level one fields
$\delta|\Phi\rangle=Q_{BRST}|\Psi\rangle$ are given by the general
level one field of ghost number $-3/2$ :
$|\Psi\rangle=\rho(Y)b_{-1}|\downarrow\rangle$ (In the Neumann case
this gives The Maxwell equation for $\omega$ if the auxiliary field
$\lambda$ is integrated out). Here the situation is entirely
different,    there are no  conditions on the wave-function at
all. This seems to be an indication that the level one open string
field is a Lagrange multiplier field and that the proper way to deal
with it is to integrate it out. This is very difficult since as we
have seen this state couples to closed string (scalar) states and it
influences the closed string spectrum. An investigation of this
question is given in \cite{mbg7} where it is indicated that the
dilaton might be removed from the theory.

\section{Cancellation of infinities}
As mentioned in section \ref{general} Polchinski has introduced a
different version of Dirichlet string theory. In this scheme  all
boundaries are mapped to the same space time point. In the
calculation of scattering amplitudes one usually takes just connected
diagrams into account (the part of the S-matrix which does not have
trivial delta functions in a subset of momenta is important). In this
scheme the disconnected world sheet can exchange momenta via their
common space time point and so they are not disconnected from a space
time point of view. The power of  the closed string coupling constant
$g_{cl}$ is given in terms of the number of diagrams $m$,    the
number of vertices $k_i$,    the number of boundaries $b_i$,    the
number of handles $h_i$ on diagram $i$,    where $i=1,    . . ,    m$
\begin{equation}
g_{cl}^{\{-2m+\sum b_i+\sum k_i+2\sum h_i \}}
\end{equation}
This changes in the way the perturbation expansion is organized has
two notable consequences,    namely that for every diagram an
arbitrary number of disks with no vertex operator insertions has to
be introduced each of which gives a contribution $(C
g_{cl}^{-1})^n/n! $ for $n$ disks. Summing over $n$ gives a factor
$\exp (C/g_{cl})$ where $C$ is a negative constant.
The high energy fixed angle behavior is again power behaved but
differs from that in our approach since there are configurations
where every vertex operator is on a separate disk,    which gives a
momentum independent $s^0$ contribution.
Polchinski has shown that for the $N=1$ divergences which we analyzed
in Section \ref{divergences} there is a cancellation between the
cases (a),    (b) and (c) if one considers diagrams of the same order
in $g_{cl}$ but of different topology.
Note that the $N=1$ state which is responsible for the divergence
enters the amplitudes in the following way
\begin{equation}
\langle V_1 . .  V_k\oint\partial_n
X_{\mu}\rangle_{\Sigma}=\sum_{j=1}^{k}k_j^{\mu}\langle V_1. .
V_k\rangle_{\Sigma}
\end{equation}
Here $V_i$ denotes a physical state vertex operator with momentum
$k_i^{\mu}$ and $\langle. .  \rangle_{\Sigma}$ denotes the functional
integral over the moduli space of $\Sigma$.
As an example we now indicate how the divergences in Figure 3 cancel
in this scheme.
The $N=1$ divergent contribution of the diagram (3.a) is given by
\[
A_{(a)}=\int\frac{d\eta}{\eta}\langle
V_1V_2V_3\oint\partial_nX^{\mu}\rangle_{\Sigma}\langle
\oint\partial_nX_{\mu}V_4\rangle_{D}
\]
\begin{equation}
=\int\frac{d\eta}{\eta}\langle V_1V_2V_3\rangle_{\Sigma}\langle
V_4\rangle_D(k_1+k_2+k_3)_{\mu}k_4^\mu
\end{equation}
To cancel this divergence we add two more diagrams which contain
divergences of type (c) in order to give an complete  square of
momenta $(\sum_{i=1}^4k_i)^2$ which vanishes upon momentum
conservation, since the single boundary position $Y$ is integrated.
The two diagrams are given by
\begin{eqnarray*}
A_{(a)}^{\prime} &=& \frac{1}{2}\int\frac{d\eta}{\eta}\langle
V_1V_2V_3\oint\partial_nX^{\mu}\oint\partial_nX_{\mu}
\rangle_{\Sigma}\langle V_4\rangle_{D} \\
 &+& \frac{1}{2}\int\frac{d\eta}{\eta}\langle
V_1V_2V_3\rangle_{\Sigma}\langle
\oint\partial_nX^{\mu}\oint\partial_nX_{\mu}V_4\rangle_{D} \\
&=& \frac{1}{2}\int\frac{d\eta}{\eta}\langle
V_1V_2V_3\rangle_{\Sigma}\langle
V_4\rangle_{D}\{k_4^2+(k_1+k_2+k_3)^2\}
\end{eqnarray*}
$A_{(a)}+A_{(a)}^{\prime}$ vanishes now because of momentum
conservation. The factor $1/2$ comes from the symmetry factor for the
second diagrams.
The divergence of diagram (3.b) is cancelled in a similar way
\begin{eqnarray*}
A_{(b)}&=&\int\frac{d\eta}{\eta}\langle
V_1V_2\oint\partial_nX^{\mu}\rangle_{\Sigma_1}
\langle\oint\partial_nX_{\mu}V_3V_4\rangle_{\Sigma_2} \\
A_{(b)}^{\prime}&=&\int\frac{d\eta}{\eta}\langle
V_1V_2\oint\partial_nX_{\mu}\oint\partial_nX^{\mu}
\rangle_{\Sigma_1}\langle V_3V_4\rangle_{\Sigma_2} \\
&+& \int\frac{d\eta}{\eta}\langle V_1V_2\rangle_{\Sigma_1}\langle
\oint\partial_nX_{\mu}\oint\partial_nX^{\mu}V_3V_4\rangle_{\Sigma_2}
\end{eqnarray*}
Note that in this cancellations it is actually not enough to just
look at planar diagrams as an example, the amplitudes
$A^{\prime}_{(a)}$ and $A^{\prime}_{(b)}$ contain the insertion of
the two boundary vertex operators at different components of the
boundary and this corresponds to a degeneration of a nonplanar
surface.
The divergence of type (c) in the case of a single surface is
actually zero,    since
\[
A_{(c)}=\int\frac{d\eta}{\eta}\langle
V_1V_2V_3V_4\oint\partial_nX^{\mu}\oint\partial_nX_{\mu}V_4
\rangle_{\Sigma}
\]
\begin{equation}
=\int\frac{d\eta}{\eta}\langle V_1V_2V_3
V_4\rangle_{\Sigma}(k_1+k_2+k_3+k_4)^2
\end{equation}
It is clear that this method can be applied to the divergences (a) to
(c) in general. This indicates that the divergence problem of the
insertion of Dirichlet boundaries might not make the theory
inconsistent but lead to some new interesting physics.

\pagebreak
\begin{figure}
\epsfxsize=400pt\epsffile{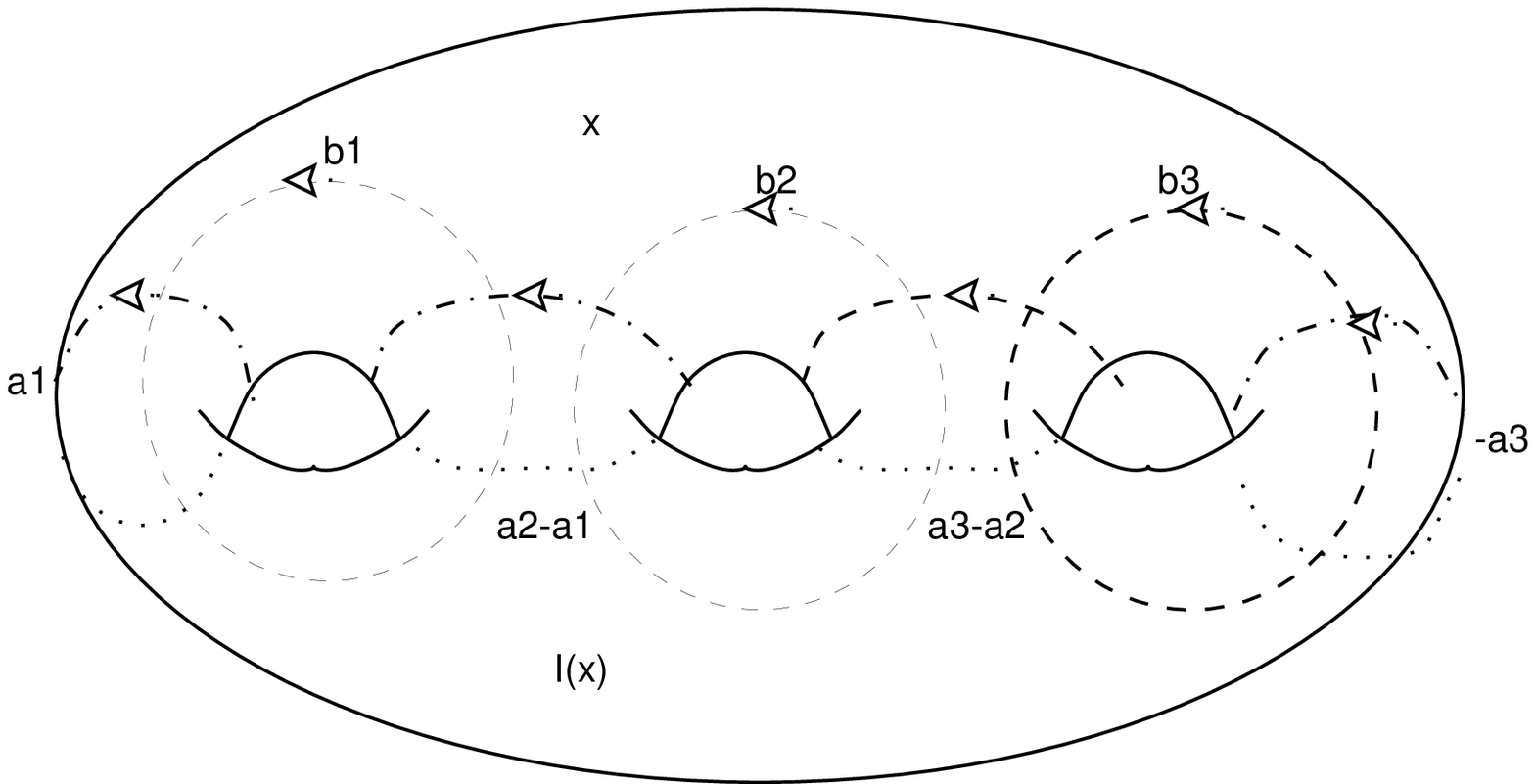}
\caption[a]{\protect\label{FIGURE1}Canonical homology cycles on a
surface of genus 3,    The involution I acts in a way that the linear
combinations of a cycles are fixed under I}
\end{figure}
\begin{figure}
\epsfxsize=300pt\epsffile{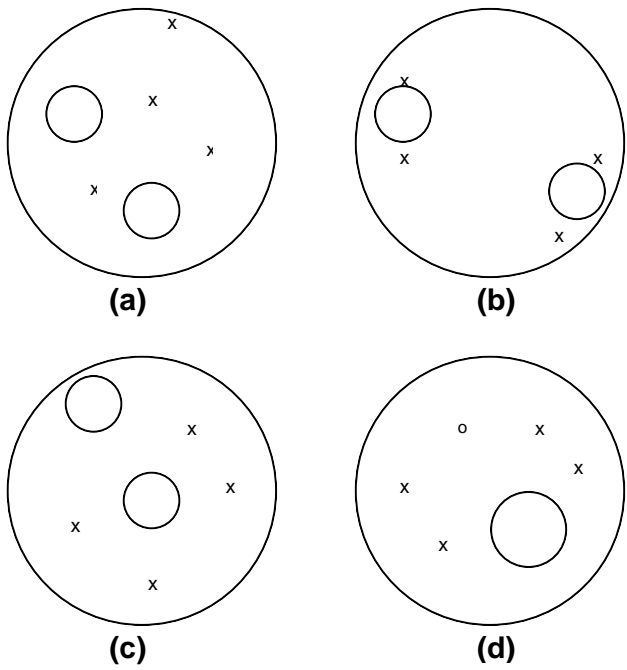}
\caption[a]{\protect\label{FIGURE2}The degenerations of an open
string with three boundaries and for vertex operators: a) one vertex
going to the boundary b) surface is separated into two parts c) two
boundaries touching d) one boundary shrinking to a point }
\end{figure}
\begin{figure}
\epsfxsize=400pt\epsffile{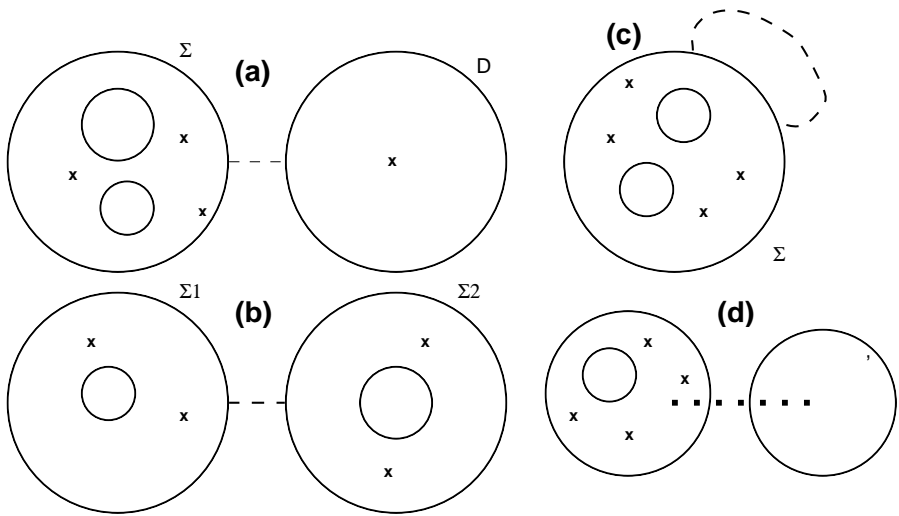}
\caption[b]{\protect\label{FIGURE3} a)-c) The degenerations
represented by the insertion of two boundary vertex operators and a
open string propagator d) the degeneration is represented by closed
string vertex operator on the surface and a closed string propagator
coupling to a disk}
\end{figure}
\begin{figure}
\epsfxsize=400pt\epsffile{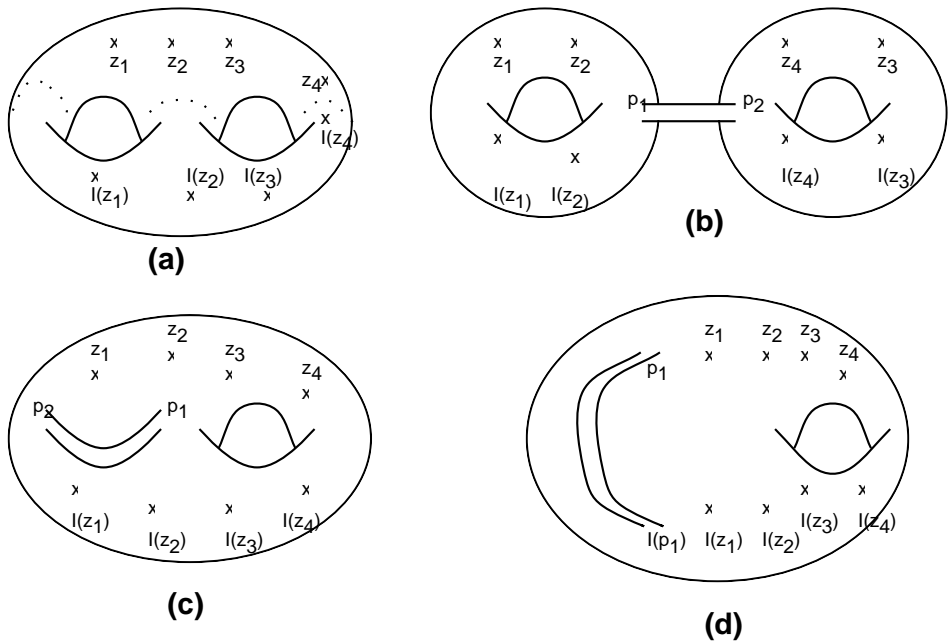}
\caption[a]{\protect\label{FIGURE4}The degenerations of open strings
represented as degenerations of the doubled surface which is
symmetric under I a) one vertex and its mirror image coming close at
the boundary b) a zero homology cycle pinching separating the surface
into disconnected parts c) a b-cycle pinching,    for b),    c) the
plumbing fixtures $p_1$ and $p_2$ are invariant under I,    d) an
a-cycle pinching}
\end{figure}
\begin{figure}
\epsfxsize=300pt\epsffile{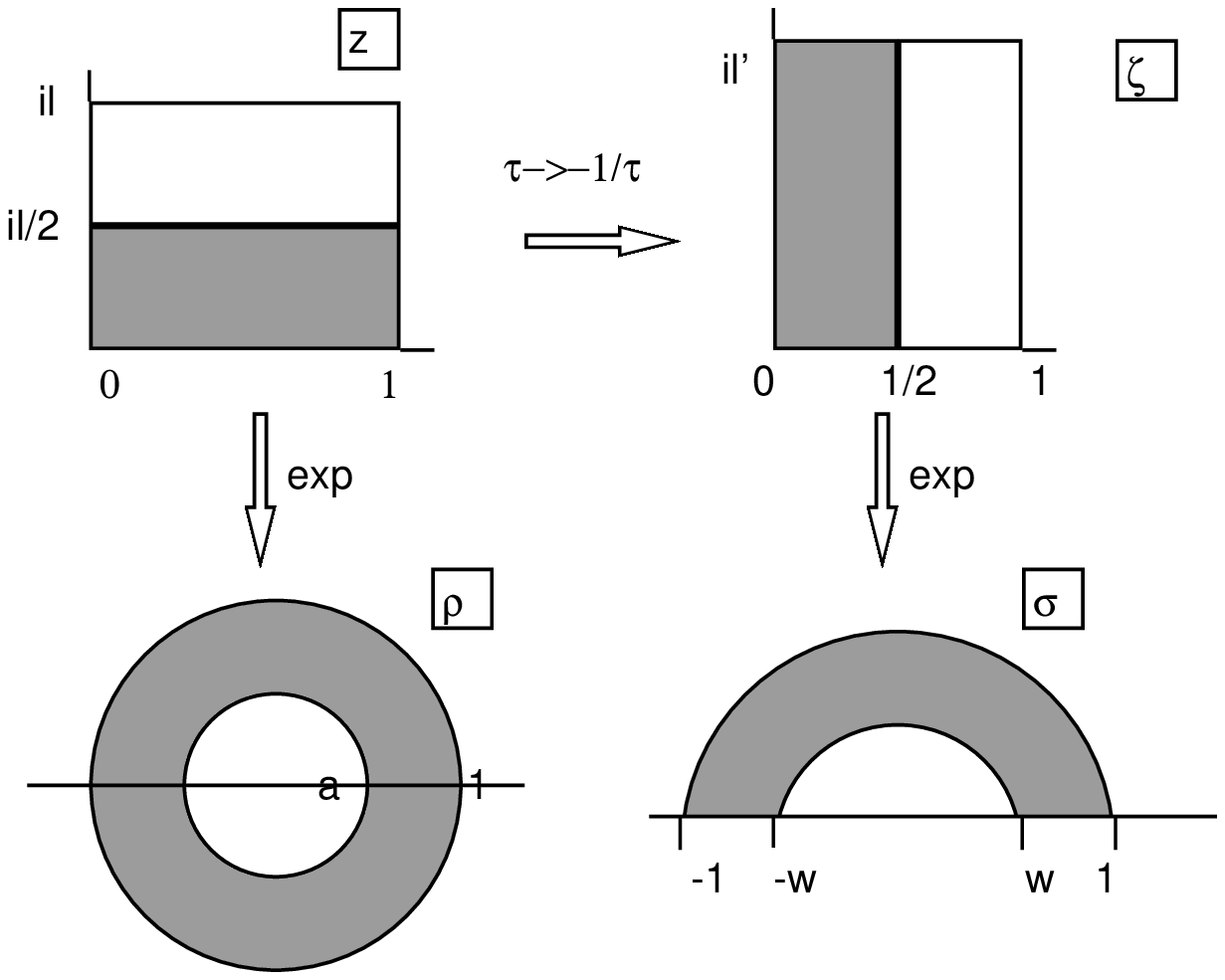}
\caption[a]{\protect\label{FIGURE5} The different variables used for
the annular world sheet. The $z$ plane represents the fundamental
cell,    the $\zeta$ plane the fundamental cell after the modular
transformation. The annulus in the $\rho$ plane and the semiannulus
in the $\sigma$ plane are given by exponential mappings. }
\end{figure}


\begin{thebibliography}{99}
\bibitem{cardy} J.L. Cardy, Nucl. Phys B324 (1989) 581
\bibitem{schwarz} J. Schwarz, Nucl. Phys. B65 (1973) 131
\bibitem{corrigan} E. Corrigan and D. Fairlie, Nucl. Phys. B91 (1975)
527
\bibitem{mbgd} M.B. Green J.A. Shapiro, Phys. Lett. B64 (1976) 454
\bibitem{cohen} A.Cohen et al., Nucl. Phys B281 (1987) 127
\bibitem{mbgtsd} M.B. Green, Phys. Lett. B266 (1991) 325
\bibitem{pol2} J. Dai,R.G. Leigh and J. Polchinski, Mod. Phys. Lett.
A4 (1989) 2073
\bibitem{li} M. Li, Nucl. phys B420 (1994) 339
\bibitem{zhang} Z.Yang preprint UR-1288/ER-40685-727 (unpublished)

\bibitem{pol3} J.Polchinski, Phys. Rev D50 (1994) 6041

\bibitem{knizh} A.A. Belavin and V.G. Knizhnik, Phys. Lett. B168
(1986) 201
\bibitem{verlinde} E. Verlinde and H. Verlinde, Nucl. Phys. B288
(1987) 357

\bibitem{shiff} M. Shiffer and D.C. Spencer, Functionals of finite
Riemann surfaces (Princeton Univ. Press, 1954)

\bibitem{burgess} C.P. Burgess and T.R. Morris, Nucl. Phys B291
(1987) 256
\bibitem{blau}  S.K. Blau et. al., Nucl. Phys. B301 (1988) 285
\bibitem{sagnotti} M. Bianchi and A.Sagnotti, Phys. Lett. B231 (1989)
389
\bibitem{dhoker} E'Dhoker and D.H. Phong, Rev. of mod. Phys. 60
(1988) 917 and references therein

\bibitem{mbg2} M.B. Green, Nucl. Phys. B124 (1977) 461
\bibitem{mbg4} M.B. Green, Phys. Lett. B282 (1992) 380
\bibitem{pol1} J.Polchinski, Phys. Rev. Lett. 68 (1992) 1267

\bibitem{veneziano} G. Veneziano, Nuov. Cim. 57a (1968) 190
\bibitem{amati} V. Allesandrini,D. Amati and B. Morel, Nuov. Cim. 7A
(1971) 797
\bibitem{gross1} D.J. Gross and P.F. Mende, Nucl. Phys. B303 (1988)
407
\bibitem{callan1} C.G. Callan et al.,Nucl. Phys. B293 (1987) 83
\bibitem{farkas} H. Farkas and I. Kra, Riemann surfaces (Springer,
Berlin, 1980)
\bibitem{Fay} J. Fay, Theta functions on Riemann Surfaces, Lect.
notes in math. 323 (Springer, Berlin 1973)

\bibitem{bateman} Erdelyi et al., Bateman manuscript (McGraw-Hill,
New York, 1955)

\bibitem{marcus} N. Marcus, Phys. Lett. B219 (1989) 265
\bibitem{Fischler} W. Fischler and L. Susskind, Phys. Lett. B171
(1986) 383
\bibitem{mbg7} M.B. Green, Phys. Lett. B302 (1993) 29
\bibitem{gsw} M.B. Green,J.H. Schwarz and E. Witten, Superstring
theory (Cambridge University Press,1987)
\bibitem{mbg8} M.B. Green, Phys. Lett. B329 (1994) 435
\bibitem{gava} E. Gava et al., Phys. Lett. B168 (1986) 207
\bibitem{pol5} J. Polchinski, Nucl. Phys. B307 (1988) 61
\bibitem{Sonoda} H. Sonoda, Nucl. Phys. B311 (1989) 401,417
\end{thebibliography}
\end{document}